\pdfoutput=1

\documentclass{article}

\usepackage{graphicx,epstopdf}
\usepackage[english]{babel}
\usepackage[fleqn]{amsmath}
\usepackage{amssymb,amsthm}
\usepackage{lmodern,microtype}
\usepackage[a4paper,left=4cm]{geometry}
\usepackage{hyperref}

\usepackage{tikz}
\usetikzlibrary{arrows,decorations}

\renewcommand{\i}{\text{i}}
\newcommand{\rrangle}{\rangle\hspace{-.05cm}\rangle}

\newcommand{\ket}[1]{|#1\rangle}

\newcommand{\beq}{\begin{equation}}
\newcommand{\eeq}{\end{equation}}

\newtheorem*{theorem}{Theorem}
\newtheorem*{conjecture}{Conjecture}

\title{\large \bf Bethe ansatz solvability and supersymmetry 
of the $M_2$ model of single fermions and pairs}
\date{}
\author{\normalsize \sc{Christian Hagendorf}$^1$, \sc{Thessa B. Fokkema}$^2$ and \sc{Liza Huijse}$^3$
\bigskip\\
{\normalsize
  \begin{minipage}{\textwidth}
  \begin{center}
  \textit{
  $^1$ Universit\'e Catholique de Louvain\\
  Institut de Recherche en Math\'ematique et Physique\\
  Chemin du Cyclotron 2, 1348 Louvain-la-Neuve, Belgium
  \medskip\\
$^2$ Institute for Theoretical Physics, University of Amsterdam\\
 Science Park 904, 1090 GL Amsterdam, The Netherlands
\medskip\\
  $^3$ Physics Department, SITP\\
382 Via Pueblo Mall,Varian Lab\\
Stanford University, Stanford CA, 94305-4060}
\bigskip\\
  \href{mailto:christian.hagendorf@uclouvain.be}{\normalsize \texttt{christian.hagendorf@uclouvain.be}},
  \href{mailto:t.b.fokkema@uva.nl}{\texttt{t.b.fokkema@uva.nl}},\\
   \href{mailto:lhuijse@stanford.edu}{\texttt{lhuijse@stanford.edu}}
    \end{center}
  \end{minipage}
}
}

\begin{document}
\maketitle
\begin{abstract}
A detailed study of a model for strongly-interacting fermions with exclusion rules and lattice $\mathcal N=2$ supersymmetry is presented. A submanifold in the space of parameters of the model where it is Bethe-ansatz solvable is identified. The relation between this manifold and the existence of additional, so-called dynamic, supersymmetries is discussed. The ground states are analysed with the help of cohomology techniques, and their exact finite-size Bethe roots are found. Moreover, through analytical and numerical studies it is argued that the model provides a lattice version of the $\mathcal N=1$ super-sine-Gordon model at a particular coupling where an additional $\mathcal N=(2,2)$ supersymmetry is present. The dynamic supersymmetry is shown to allow an exact determination of the gap scaling function of the model.
\end{abstract}


\section{Introduction}

In recent years there has been increasing interest in the study of supersymmetry in the context of condensed matter and statistical mechanics systems. By now a range of systems have been identified, including experimentally realisable and physically relevant systems, which exhibit supersymmetry. Examples include the spin-$1/2$ XXZ chain at anisotropy $\Delta=-1/2$ \cite{fendley:03,VW}, topological superconductors at a phase transition \cite{grover:14} and polar molecules in an optical lattice at a multicritical point \cite{HuijseBauerBerg14}. 

Much progress in understanding the value and significance of studying supersymmetry in condensed matter systems was made with and following the introduction of a quantum mechanical model for lattice fermions with explicit $\mathcal{N}=2$ supersymmetry \cite{fendley:03_2}. It was found that many of the special features of supersymmetric theories carry over to these lattice models. In particular, the special properties of the ground states of supersymmetric theories make these lattice models amenable to analytical studies even in the strongly interacting regime. These features were used to reveal that on graphs of spatial dimensions $d>1$ these models typically exhibit a strong form of quantum charge frustration, called superfrustration, which is characterised by an extensive ground state entropy \cite{fendley:05_2,HvE,JJ,Tri12}. 

In the supersymmetric model the fermions obey a hard-core constraint such that all sites neighbouring an occupied site have to be empty. This model is the first member ($k=1$) of the supersymmetric $M_k$ models for fermions which are subject to the constraint that there are at most $k$ particles in connected particle clusters. Studying these supersymmetric models for lattice fermions on $d=1$ dimensional chains has also led to a number of interesting discoveries, most remarkably, the realisation that the spin-$1/2$ XXZ chain at anisotropy $\Delta=-1/2$ possesses a hidden supersymmetry on the lattice \cite{fendley:03,VW, hagendorf:13}. This follows from an intricate mapping between the supersymmetric model and the spin chain. A similar mapping allows to relate the $M_2$ model to the integrable spin$-1$ XXZ chain, with anisotropy tuned to a value where it also possesses a hidden supersymmetry on the lattice. In \cite{hagendorf:13, meidinger:13} similar hidden supersymmetries were studied in more detail and named dynamic supersymmetries. Since the mapping between the $M_{1,2}$ models and the spin models is quite intricate, it is not obvious that the supersymmetric fermion models have in general to be integrable as well. However, it turns out that they are indeed solvable for specific choices of their coupling constants via a coordinate Bethe Ansatz \cite{fendley:03}.

Integrability is arguably one of the most powerful analytical tools available for the study of $d=2$ classical or $d=1$ quantum mechanical systems \cite{baxterbook}. Integrable models are very special and enjoy a high degree of analytical structure. The integrable supersymmetric models provide an interesting opportunity to study the interplay between the structure due to integrability and that due to supersymmetry (see also \cite{meidinger:13}). This is the topic of the present work.

In this paper we study an inhomogeneous version of the $M_2$ model with interactions which are site-dependent. For a chain of length $N$ these inhomogeneous interactions can be parameterised by $2N$ parameters: $\lambda_x$ and $\mu_x$, where $x=1,\dots,N$ labels the sites of the chain. For this model we identify a special two-parameter submanifold in the $(\lambda_x,\mu_x)$-parameter space where the model possesses two types of dynamic supersymmetry in addition to the explicit $\mathcal{N}=2$ supersymmetry. Furthermore, from completely independent considerations we find that the model is solvable via a generalised coordinate Bethe ansatz \cite{blom:12} on precisely this same special submanifold in parameter space. 

\begin{figure}[h]
\centering
\includegraphics[width=0.65\columnwidth]{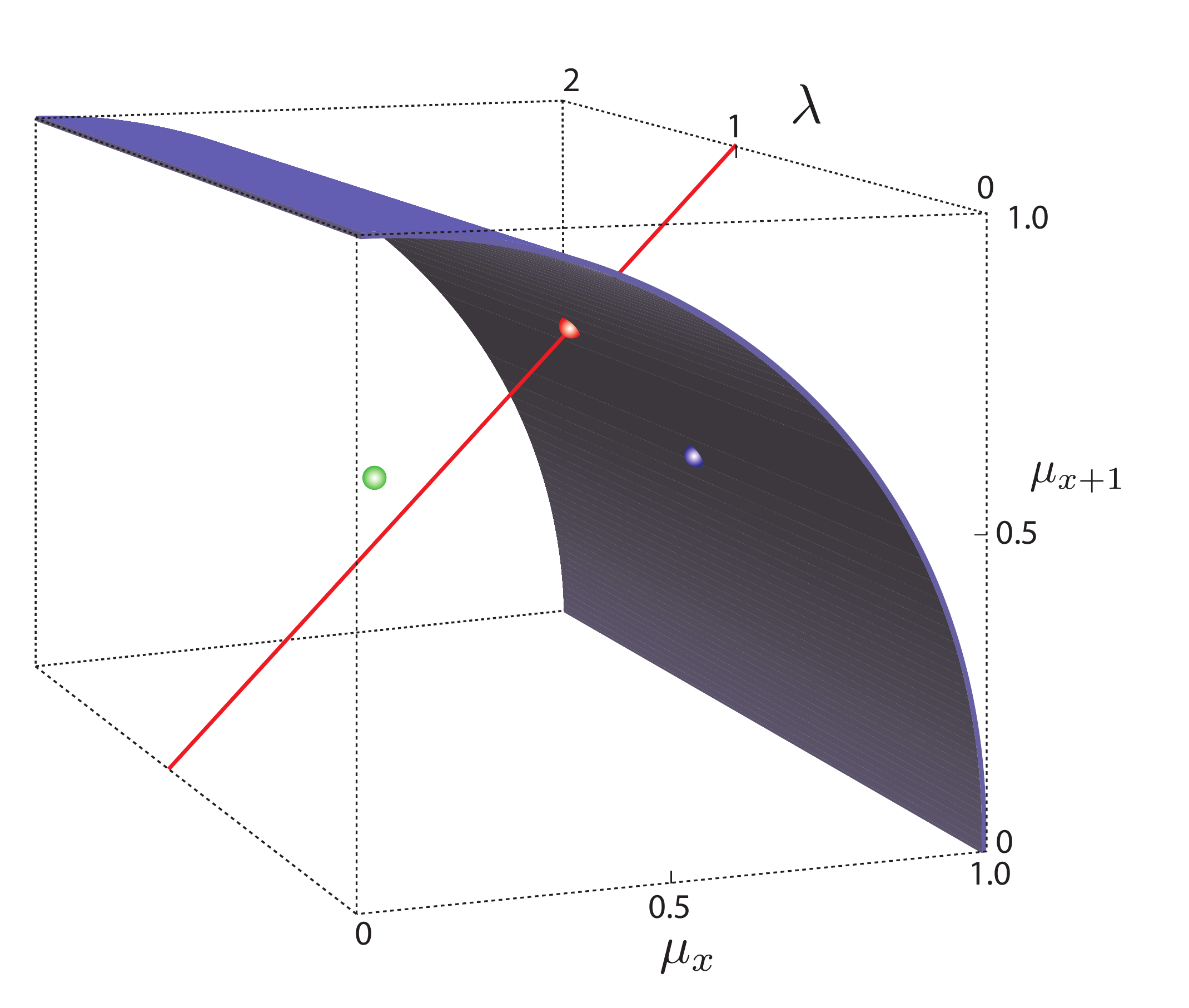}
\caption{We plot the special two-parameter submanifold (\ref{eqn:integrableline}), where the model is integrable and enjoys additional supersymmetry, in a three-parameter subspace for which $\lambda_{x+2}=\lambda_{x}$, $\mu_{x+2}=\mu_x$ for all $x$. We define $\lambda=\lambda_{x}/\lambda_{x+1}$ and plot the range $0<\lambda<2$ and $\mu_x>0$. The red line indicates the parameters for which the original $M_2$ model is recovered: $\lambda=1$ and $\mu_x=\mu_{x+1}=\mu$. The red point lies at the intersection of the red line with the special submanifold: $\mu=1/\sqrt{2}$. It is the point where the original model is integrable and maps to the spin-1 chain. The blue point is a generic point on the special submanifold, and the green point is a generic point that is neither on the homogeneous line nor on the special submanifold. In section \ref{sec:gap} we will compare numerical data for these three points. Note that the model possesses the explicit $\mathcal{N}=2$ supersymmetry for general $\lambda_x$ and $\mu_x$ and thus, in particular, everywhere in the plotted parameter space.  \label{fig:param}}
\end{figure}

The special submanifold is given by
\begin{equation}
\lambda_{x+2}=\lambda_{x}, \quad \mu_{x}^2+\mu_{x+1}^2=1,  \label{eqn:integrableline}
\end{equation}
and is depicted in figure \ref{fig:param}. The homogeneous model, i.e. the original $M_2$ model with site-independent interactions, intersects the special submanifold at the point: $\lambda_x=1$ and $\mu_x=\mu=1/\sqrt{2}$ for all $x$. These are indeed precisely the parameters for which the original $M_2$ model was found to be integrable \cite{fendley:03}. In the original model the constraint on $\mu$ follows from spin-reversal symmetry upon mapping the model to the spin$-1$ chain. Here we will see that the constraint $\mu_{x}^2+\mu_{x+1}^2=1$ is also related to a type of spin-reversal symmetry, even though for site-dependent interactions a mapping to a spin model is not known at present.

We investigate how the supersymmetry translates into the structure of the Bethe ansatz. We find, in particular, a relation between the action of the supercharges, the generators of the supersymmetry, and exact complete strings of Bethe roots. Furthermore, we conjecture that a particular zero energy state, which is a supersymmetry singlet, is a Bethe wave function whose Bethe roots all take the same value.

The paper is organised as follows. We define the model and present a detailed analysis of its various symmetries, including the dynamic supersymmetries upon restricting to the special submanifold in section \ref{sec:overview}. In section \ref{sec:ba} we study the model by means of a generalised coordinate Bethe ansatz solution: as we shall see, the requirement that the model be Bethe-ansatz solvable leads again to a restriction to the special submanifold. This is followed by an investigation of the connections between supersymmetry and the Bethe ansatz in section \ref{sec:bethesusy}. The ground states of the model are analysed in section \ref{sec:cohomology}. We compute the number of zero-energy states of the $M_2$ model using cohomology techniques. This is a natural extension to the Witten index, which gives a lower bound on the number of ground states, and was computed in \cite{fendley:03}. Moreover, we find the exact finite-size Bethe root distribution of the ground states from a system of functional equations. In section \ref{sec:gap} we briefly discuss the field theory interpretation of the special submanifold in terms of a particular supersymmetry-preserving perturbation of the second superconformal minimal model, which describes the continuum theory of the homogeneous $M_2$ model with $\mu=1/\sqrt{2}$. We then discuss the consequences of the dynamic supersymmetry on the gap scaling away from the homogeneous point and find that it allows us to determine the shape of the scaling function. We present our conclusions in section \ref{sec:conclusion}.


\section{The model and its symmetries}\label{sec:overview}
In this section we introduce the inhomogeneous $M_2$ model and discuss its symmetries. In particular, we identify a special submanifold in parameter space where the model has two types of dynamic supersymmetry in addition to the original $\mathcal N=2$ supersymmetry.

\subsection{Definition of the model}
\label{sec:defmodel}
The $M_2$ model is the second member of a series of lattice fermion models with supersymmetry introduced by Fendley, Schoutens and Nienhuis \cite{fendley:03}. It describes spinless fermions on a lattice subject to the exclusion constraint that connected particle clusters may not contain more than two fermions. Hence, any allowed configuration consists of isolated fermions without neighbours or fermion pairs on adjacent sites.

\paragraph{Hilbert space.} While the model can be defined on arbitrary graphs, in this work we restrict our considerations to the one-dimensional chain with $N$ lattice sites, labeled by integers $x=1,\dots,N$. The chain is closed, which means that the sites $N+1$ and $1$ are identified. We visualise fermion configurations on these closed chains by empty sites $\circ$, and occupied sites $\bullet$. For instance, an admissible configuration for $N=6$ sites and $f=3$ fermions is represented by $\bullet \circ \bullet \circ \circ \hspace{.1cm} \bullet$. It contains an isolated particle at $x=3$, and a pair on the sites $x=6,1$. Conversely, an example for a non-admissible configuration is given by $\bullet \bullet \circ \circ \circ \hspace{.1cm}\bullet$ because it violates the exclusion constraint: there are three consecutive occupied sites $x=6,1,2$.

The Hilbert space $\mathcal H_N$ of the model is a standard fermionic Fock space. Its canonical basis is given by a set of mutually orthonormal vectors labeled by all admissible fermion configurations, for example $|{\bullet \circ \bullet \circ \circ \,\bullet}\rangle$. The Hilbert space decomposes naturally into subsectors of constant fermion number, i.e. eigenspaces $\mathcal H_{N,f}$ of the fermion number operator $F$.

\paragraph{Supercharge.} We define a supercharge $Q_+:\mathcal H_{N,f} \to \mathcal H_{N,f-1}$ expanding slightly on the construction of \cite{fendley:03} in order to arrive at the inhomogeneous $M_2$ model. First, we define the fermion annihilation operators $Q_{+,x}$, which take out a fermion from site $x$ with an amplitude $\lambda_{1,1,x}$, $\lambda_{2,1,x}$ or $\lambda_{2,2,x}$, depending on whether the fermion is isolated or first/second member of a pair. Throughout this work we choose the amplitudes $\lambda_{a,b,x}$ to be real positive. The action on simple basis vectors is therefore given by
\begin{align*}
  & Q_{+,x}|\cdots \underset{x}{\circ}\cdots \rangle = 0,\\
  & Q_{+,x}|\cdots \circ \underset{x}{\bullet}\circ \cdots \rangle = (\pm)\lambda_{1,1,x}|\cdots \circ \underset{x}
 {\circ}\circ \cdots \rangle,\\
  & Q_{+,x}|\cdots \circ \underset{x}{\bullet}\bullet\circ \cdots \rangle = (\pm)\lambda_{2,1,x}|\cdots \circ \underset{x}
 {\circ}\bullet\circ \cdots \rangle,\\
 & Q_{+,x}|\cdots \circ \bullet \underset{x}{\bullet}\circ \cdots \rangle = (\pm)\lambda_{2,2,x}|\cdots \circ\bullet \underset{x}
 {\circ}\circ \cdots \rangle,
\end{align*}
where the sign $(\pm)$ is the usual fermionic string, i.e. minus one to the number of fermions located to the left of the site $x$. The supercharge of the model is defined as the sum of these operators over all sites
\begin{equation}
  Q_+ = \sum_{x=1}^N Q_{+,x}.
  \label{eqn:Qplus}
\end{equation}
Throughout this article, we assume that the weights are real positive numbers. The original $M_2$ model of \cite{fendley:03} is recovered by setting $\lambda_{a,b,x}=\lambda_{a,b}$ for all $x$. The condition $(Q_+)^2=0$ imposes the constraint
\begin{equation}
  \lambda_{1,1,x+1}\lambda_{2,1,x} = \lambda_{1,1,x}\lambda_{2,2,x+1}
  \label{eqn:lambdaconstraint}
\end{equation}
on the staggering constants. This relation is conveniently solved by the following choice of parameters
\begin{equation}
  \lambda_{1,1,x} = \lambda_x,\quad \lambda_{2,1,x} = \lambda_x \mu_x, \quad \lambda_{2,2,x} = \lambda_x\mu_{x-1}.
  \label{eqn:choicelambda}
\end{equation}
\paragraph{Hamiltonian.} The supercharge and its Hermitian conjugate $\bar Q_+ = (Q_+)^\dagger$ generate the Hamiltonian of the model $H=\{Q_+,\bar Q_+\}$ which preserves the fermion number. Instead of writing it out in terms of fermion creation and annihilation operators we provide a list with the non-vanishing amplitudes which describe elementary hopping processes and interactions between adjacent particle clusters. The hopping terms are
\begin{subequations}
\begin{itemize}
  \item \textbf{Single hop}
  \begin{equation}
    \circ\hspace{-.05cm}\underset{x}{\bullet}\hspace{-.05cm}\circ\circ\quad \leftrightarrow \quad \circ\hspace{-.05cm}\underset{x}{\circ}\hspace{-.05cm}\bullet\hspace{-.05cm} \circ \qquad \text{with amplitude}\quad\lambda_x\lambda_{x+1}(1-\mu_x^2),
    \label{eqn:singlehop}
  \end{equation}
  \item \textbf{Pair hop}
  \begin{equation}
    \circ\underset{x}{\bullet}\bullet\circ\circ\quad \leftrightarrow \quad \circ\underset{x}{\circ}\bullet\bullet\circ\qquad \text{with amplitude}\quad-\lambda_x\lambda_{x+2}\mu_x\mu_{x+1},
  \end{equation}
  \item \textbf{Split-join on the right and left}
  \begin{align}
    \circ\underset{x}{\bullet}\circ\bullet\circ\quad \leftrightarrow \quad \circ\underset{x}{\bullet}\bullet\circ\circ\qquad \text{with amplitude}\quad \lambda_{x+1}\lambda_{x+2}\mu_x,\\
    \circ\underset{x}{\bullet}\circ\bullet\circ\quad \leftrightarrow \quad \circ\underset{x}{\circ}\bullet\bullet\circ\qquad \text{with amplitude}\quad\lambda_{x}\lambda_{x+1}\mu_{x+1},
  \end{align}
  \item \textbf{Partner swap}
  \begin{equation}
    \circ\underset{x}{\bullet}\bullet\circ\bullet\circ\quad \leftrightarrow \quad \circ\underset{x}{\bullet}\circ\bullet\bullet\circ\qquad \text{with amplitude}\quad\lambda_{x+1}\lambda_{x+2}\mu_x\mu_{x+2}.
    \label{eqn:partnerswap}
  \end{equation}
\end{itemize}
The potential energy is easily written in closed form
\begin{equation}
  \sum_{x=1}^N \lambda_x^2\left(P_{x-1}P_{x+1} + \mu_{x-1}^2P_{x-2}(1-P_{x-1})P_{x+1}+\mu_{x}^2P_{x-1}(1-P_{x+1})P_{x+2}\right)
\end{equation}
where $P_{x}=1-n_x$ and $n_x$ is the local occupation number operator acting on site $x$. The contribution of the site $x$ to the potential energy of a given configuration is:
\begin{itemize}
  \item $\lambda_x^2$ if it is possible to create or annihilate an isolated particle at site $x$,
  \item $\lambda_x^2\mu_{x-1}^2$ if it is possible to create or annihilate a particle at site $x$ which is part of a pair on sites $x-1,x$,
  \item $\lambda_x^2\mu_{x}^2$ if it is possible to create or annihilate a particle at site $x$ which is part of a pair on sites $x,x+1$.
\end{itemize}
\end{subequations}

The supersymmetry implies, in particular, that $H$ is positive definite (its eigenvalues are positive or zero). All solutions to the Schr\"odinger equation $H|\psi\rangle = E|\psi\rangle$ with strictly positive energy $E>0$ organise into doublets $(|\psi\rangle, Q_+|\psi\rangle)$, where $\bar Q_+|\psi\rangle = 0$. The states $|\psi \rangle$ and $ Q_+|\psi\rangle$ are called superpartners. Possible zero-energy states, if they exist, are singlet representations of the supersymmetry algebra: they solve the equations $Q_+|\psi\rangle = \bar Q_+|\psi\rangle = 0$.

\paragraph{Periodic staggering and translation symmetry.} When all the parameters of the model are site-independent, the Hamiltonian and the supercharges commute with the standard translation operator $T$ on the fermionic Fock space. A controlled way of breaking this invariance under translation is to introduce a periodic modulation of the parameters $\lambda_{x+p}=\lambda_x,\,\mu_{x+p}=\mu_x$ for some integer $p>1$. This requires of course that the length of the chain is an integer multiple of $p$, and then leads to the invariance $[H,T^p]=[Q_+,T^p]=[\bar Q_+,T^p]=0$. The interest of such a modulation is the following: it was observed in \cite{fendley:10,blom:12} that for the closely related $M_1$ model a well-chosen period $p$ allows to drive the model off criticality while retaining its integrability.

Other possible generalisations are so-called twisted boundary conditions. They amount to a modification of the hopping amplitudes nearby the sites $x=N,1$: whenever a particle hops from site $N$ to $1$, the hopping amplitude is multiplied by a factor $e^{\i \phi}$, whereas for the reverse process it is multiplied by $e^{-\i\phi}$. Here, $\phi$ is the so-called twist angle. The appropriate translation operator for this case is $T'= T e^{\i \phi (n_N -F/N)}$, where $n_N$ is the fermion number operator of the last site. If, additionally, the model is staggered with period $p$, we have thus $[H,(T')^p]=0$ where it is understood that the hopping amplitudes of $H$ are modified as just explained.

In this case, however, the Hamiltonian no longer possesses the supersymmetric structure defined above. The reason for this is the following. For $\phi\neq 0$ the eigenstates of the Hamiltonian are eigenstates of $(T')^p$. A supersymmetry of the type \eqref{eqn:Qplus} would therefore be a mapping within the eigenspaces of this operator. However, the supercharge $Q_+$ cannot be consistently defined for $\phi\neq 0$ as such a mapping. Indeed $Q_+$ is to be invariant under the twisted translation operator by $p$ sites, i.e. $(T')^p Q_+ (T')^{-p} = Q_+$, then we obtain $\lambda_{a,b,x+p} = \lambda_{a,b,x}$ for $x=1,\dots,N-p$, and $e^{\i (N-1)\phi/N}\lambda_{a,b,N-j}=\lambda_{a,b,p-j},\, j=0,\dots, p-1$, by using the definition of the local operators $Q_{+,x}$. The resulting equations for the coupling constants have however a non-trivial solution only for $\phi =0$, what proves our statement.

\subsection{Symmetry enhancement: additional supersymmetries}
\label{sec:susy}
The purpose of this section is to give an overview of the symmetries of the $M_2$ model. We start with a hidden $\mathbb Z_2$ symmetry in section \ref{sec:spinreversal}. In the sections \ref{sec:dynsusy} and \ref{sec:dynneutralsusy} that the model possesses two further hidden $\mathcal N=2$ supersymmetries on the line of couplings \eqref{eqn:integrableline}. For clarity we suppress some of the technical, but straightforward calculations in the presentation below.

\subsubsection{``Spin-reversal'' symmetry}
\label{sec:spinreversal}
\paragraph{Hidden spin-reversal for the homogeneous chain.} Fendley, Nienhuis and Schoutens argued that in the case of site-independent parameters $\lambda_{a,b,x}\equiv \lambda_{a,b}$ the $M_2$-model Hamiltonian is similar to the Hamiltonian of the so-called Fateev-Zamolodchikov or spin-1 XXZ chain whose Hamiltonian is of the form \cite{zamolodchikov:81,fateev:81}
\begin{equation}
  H_{FZ} = \sum_{x=1}^N\left( \sum_{a=1}^3 J_{a} (S_x^a S_{x+1}^a + 2(S^a_x)^2)-\sum_{a,b=1}^3 A_{ab} S_x^aS_x^bS_{x+1}^aS_{x+1}^b\right)\, .
  \label{eqn:fzhamiltonian}
\end{equation}
Here $S^1,S^2,S^3$ are the standard $\mathfrak{su}(2)$ generators in the spin$-1$ representation, and $J_a,\,A_{ab}$ are constants with $A_{aa}=J_{a}$ and $A_{ab}=A_{ba}$. The case of the spin$-1$ XXZ corresponds to the choice $J_1=J_2=1,\,J_3 = \cos 2\theta$, and $A_{12}=1,\,A_{13}=A_{23}=2\cos \theta -1 $ where $\theta$ parametrises the anisotropy of the model.

The identification between fermion clusters and spin components, which we label by $\uparrow,0,\downarrow$ goes as follows
\begin{equation*}
  \begin{tikzpicture}[>=stealth,baseline=0]
    \begin{scope}
    \clip (0,-.25) rectangle (.35,.25);
    \draw (0,0)--(.35,0);
    \draw[fill=white] (0,0) circle (2pt);
    \draw[fill=white] (0.35,0) circle (2pt);
   \end{scope}
   \draw[<->] (.85,0)--(1.35,0);
   \draw (1.85,0) node {$\downarrow,$};
   
    \begin{scope}
    \clip (3,-.25) rectangle (3.7,.25);
    \draw (3,0)--(4,0);
    \draw[fill=white] (3,0) circle (2pt);
    \draw[fill=black] (3.35,0) circle (2pt);
    \draw[fill=white] (3.7,0) circle (2pt);
   \end{scope}
   \draw[<->] (4.2,0)--(4.7,0);
   \draw (5.2,0) node {$0,$};
   
   \begin{scope}
    \clip (6.5,-.25) rectangle (7.55,.25);
    \draw (6.5,0)--(8,0);
    \draw[fill=white] (6.5,0) circle (2pt);
    \draw[fill=black] (6.85,0) circle (2pt);
    \draw[fill=black] (7.2,0) circle (2pt);
    \draw[fill=white] (7.55,0) circle (2pt);
   \end{scope}
   \draw[<->] (8.05,0)--(8.55,0);
   \draw (9.05,0) node {$\uparrow.$};
  \end{tikzpicture}
\end{equation*}
With this local mapping one can easily show that the same type of hopping processes and potential terms can be found in both the homogeneous $M_2$-model Hamiltonian and the spin-chain Hamiltonian. The amplitudes for the latter are however spin-reversal invariant. This implies that a mapping between the models can only exist if we impose a ``spin-reversal'' symmetry onto the amplitudes of the $M_2$ model.

Let us illustrate this with an example. In the spin chain, the amplitudes for the processes \begin{equation*}
  0\downarrow\quad \leftrightarrow\quad \downarrow 0, \quad \text{and} \quad 0\uparrow\quad \leftrightarrow\quad \uparrow 0
\end{equation*}  
coincide, and take the value $1$. If we impose the corresponding amplitudes for a single hop \eqref{eqn:singlehop} and a partner exchange \eqref{eqn:partnerswap} in the fermion model to be equal, we obtain the equation $\lambda^2(1-\mu^2)= \lambda^2 \mu^2$ which fixes $\mu = 1/\sqrt{2}$ (recall that we assume the parameters of the model to be real positive, and therefore $\mu>0$). Equality to the spin-chain amplitude leads to $\lambda=\sqrt{2}$. Quite remarkably, all other hopping amplitudes coincide for the two models provided that we set 
\begin{equation*}
  \theta = \pi/4.
\end{equation*}
The same holds for the potential energy up to an overall additive constant.

\paragraph{Extension to the inhomogeneous chain.} Given this remarkable coincidence it is natural to ask if a similar mapping to a spin model exists for the staggered $M_2$ model, and -- if so -- which type of constraint a putative spin-reversal symmetry would impose on the parameters of the model. Unfortunately, such a spin model has not been identified yet. The site-dependence of the staggering parameters excludes a translation-invariant Hamiltonian like \eqref{eqn:fzhamiltonian}. However, even without knowing the fine structure of such a model, we are free to assume that it has spin-reversal invariance, and impose it via the mapping onto the $M_2$ model.

With this idea in mind let us inspect the example considered above: we attempt to compare the amplitudes for the single hop \eqref{eqn:singlehop} and partner swap \eqref{eqn:partnerswap} processes. As both amplitudes are position-dependent we need a rule to fix their positions on the lattice. We impose the following: the single particle which changes its position hops between the same sites in both processes. For the example at hand, we impose that the following amplitudes are equal:
\begin{equation*}
  \begin{tikzpicture}
     \draw (0,0) node {$\cdots \circ\underset{x}{\bullet}\circ\circ\cdots \quad \leftrightarrow \quad \cdots \circ\underset{x}{\circ}\bullet\circ\cdots $};
     \draw (6,0) node{$\cdots \circ{\bullet}\underset{x}\circ\bullet\bullet\circ\cdots\quad \leftrightarrow \quad \cdots \circ {\bullet}\underset{x}\bullet\circ\bullet\circ\cdots$};
     \draw (0,-1) node {$\lambda_x \lambda_{x+1}(1-\mu_x^2)$};
     \draw (3,-1) node {$=$};
     \draw (6,-1) node {$\lambda_{x}\lambda_{x+1}\mu_{x-1}\mu_{x+1}$};
  \end{tikzpicture}
\end{equation*}
We see that this implies $\mu_{x+1}\mu_{x-1} = 1-\mu_x^2$. Applying this strategy to the left- and right-split-join processes, that is equating the amplitudes for the processes
\begin{equation*}
  \begin{tikzpicture}
     \draw (0,0) node {$\cdots \circ\underset{x}{\bullet}\circ \bullet \circ\cdots \quad \leftrightarrow \quad \cdots \circ\underset{x}{\circ}\bullet\bullet\circ\cdots $};
     \draw (6,0) node{$\cdots \circ\bullet\underset{x}{\circ}\bullet\circ\cdots\quad \leftrightarrow \quad \cdots \circ \bullet\underset{x}{\bullet}\circ\circ\cdots$};
     \draw (0,-1) node {$\lambda_x \lambda_{x+1} \mu_{x+1}$};
     \draw (3,-1) node {$=$};
     \draw (6,-1) node {$\lambda_{x}\lambda_{x+1}\mu_{x-1}$};
  \end{tikzpicture}
\end{equation*}
leads to $\mu_{x-1} = \mu_{x+1}$, and hence we recover the first equation of \eqref{eqn:integrableline}. It should be emphasised that while it leads to a mod$-2$ staggering for the $\mu_x$, the present argument does not impose a constraint on the $\lambda_x$.

\subsubsection{Dynamic supersymmetry (1)}
\label{sec:dynsusy}
The discussion of the last section leads naturally to the question if the ``spin-reversal'' transformation can be applied to the supercharge $Q_+$ in order to obtain a second copy of the $\mathcal N=2$ supersymmetry algebra. Let us illustrate this idea with an example. We consider the local action of $Q_{+}$ on an empty site, $x$, with empty neighbouring sites. We map this configuration and its image to spin configurations, perform a spin reversal transformation and map the result back to fermion configurations: 
\begin{equation*}
  \begin{tikzpicture}[>=stealth]

    \begin{scope}
    \clip (0,-.25) rectangle (0.7,.25);
    \draw (0,0)--(1,0);
    \draw[fill=white] (0,0) circle (2pt);
    \draw[fill=black] (0.35,0) circle (2pt);
    \draw[fill=white] (0.7,0) circle (2pt);
   \end{scope}
    \node [below] at (.35,-0.05) {$x$};

   \draw [->] (.35, -.5)--(.35,-1);
   \node [right] at (.35,-.75) {$Q_{+,x}$};

  \begin{scope}
    \clip (0,-1.75) rectangle (0.7,-1.25);
    \draw (0,-1.5)--(1,-1.5);
    \draw[fill=white] (0,-1.5) circle (2pt);
    \draw[fill=white] (0.35,-1.5) circle (2pt);
    \draw[fill=white] (0.7,-1.5) circle (2pt);
   \end{scope}
    \node [below] at (.35,-1.55) {$x$};


   \draw [->, thick] (1.9,0) -- (3.1,0);
   \node [above] at (2.5,-.75) {fermions};
   \node [below] at (2.5,-.75) {to spins};
   \draw [->, thick] (1.9,-1.5) -- (3.1,-1.5);


   \draw (4,0) node {$0$};
   \draw (4,-1.5) node {$\downarrow \downarrow$};


   \draw [->, thick] (4.9,0) -- (6.1,0);
   \node [above] at (5.5,-.75) {spin};
   \node [below] at (5.5,-.75) {reversal};
   \draw [->, thick] (4.9,-1.5) -- (6.1,-1.5);


   \draw (7,0) node {$0$};
   \draw (7,-1.5) node {$\uparrow \uparrow$};


   \draw [->, thick] (7.9,0) -- (9.1,0);
   \node [above] at (8.5,-.75) {spins to};
   \node [below] at (8.5,-.75) {fermions};
   \draw [->, thick] (7.9,-1.5) -- (9.1,-1.5);


    \begin{scope}
    \clip (10.7,-.25) rectangle (11.4,.25);
    \draw (10.7,0)--(11.4,0);
    \draw[fill=white] (10.7,0) circle (2pt);
    \draw[fill=black] (11.05,0) circle (2pt);
    \draw[fill=white] (11.4,0) circle (2pt);
   \end{scope}

   \draw [->] (11.05, -.5)--(11.05,-1);
   \node [right] at (11.05,-.75) {$Q_{-}?$};

  \begin{scope}
    \clip (10,-1.75) rectangle (12.1,-1.25);
    \draw (10,-1.5)--(12.1,-1.5);
    \draw[fill=white] (10,-1.5) circle (2pt);
    \draw[fill=black] (10.35,-1.5) circle (2pt);
    \draw[fill=black] (10.7,-1.5) circle (2pt);
    \draw[fill=white] (11.05,-1.5) circle (2pt);
    \draw[fill=black] (11.4,-1.5) circle (2pt);
    \draw[fill=black] (11.75,-1.5) circle (2pt);
    \draw[fill=white] (12.1,-1.5) circle (2pt);
   \end{scope}

  \end{tikzpicture}
\end{equation*}
The last column suggests that there might be a dynamic supercharge, which we called $Q_-$, that inserts 3 fermions and 4 sites. We show in this section that this supercharge indeed exists, provided that the staggering parameters are periodic with period two, and the model is restricted to subsectors of the Hilbert space which are invariant under translations by four sites.

\paragraph{Definition and properties.} As suggested by our example, the dynamic supercharge inserts four consecutive sites with three particles into the system: $Q_- : \mathcal{H}_{N,f} \to \mathcal{H}_{N+4,f+3}$. It is given as a linear superposition of locally acting operators $Q_{-,x}$ which perform the insertion process between sites $x$ and $x+1$:
\begin{equation*}
  Q_- = \sqrt{\frac{N}{N+4}}\sum_{x=-3}^N Q_{-,x}.
\end{equation*}
The terms with $x=-3,-2,\dots,0$ take care of the insertion process near the ``boundary'', i.e. near sites $N$ and $1$, and need to be present as the four new sites may have $N+4$ different locations on a chain with $N+4$ sites. We ignore them for the moment, and define the operators $Q_{-,x}$ for $x=1,\dots,N$. The application of the spin-reversal rules to the elementary processes induced by the supercharge $Q_+$ leads to the following actions on basis vectors:
\begin{align}
  & Q_{-,x}|\cdots \underset{x}{\textcolor{gray}{\circ}}\hspace{.08cm}\textcolor{gray}{\circ}  \cdots \rangle  = (\pm)\left(a_x|\cdots  \underset{x}{\textcolor{gray}{\circ}} \bullet \bullet\circ \bullet \hspace{.1cm}\textcolor{gray}{\circ}   \cdots \rangle +b_x|\cdots  \underset{x}{\textcolor{gray}{\circ}}\bullet \circ \bullet \bullet  \hspace{.08cm}\textcolor{gray}{\circ}  \cdots \rangle \right), \nonumber \\
  & Q_{-,x}|\cdots \underset{x}{\textcolor{gray}{\circ}}\bullet \textcolor{gray}{\circ}  \cdots \rangle  = (\pm)\,c_x|\cdots  \underset{x}{\textcolor{gray}{\circ}} \bullet \bullet \circ \bullet \bullet \textcolor{gray}{\circ}   \cdots \rangle \label{eqn:localaction},\\
  & Q_{-,x}|\cdots \underset{x}{\textcolor{gray}{\circ}}\bullet\bullet \hspace{.08cm}{\textcolor{gray}{\circ}} \cdots \rangle  = 0. \nonumber
\end{align}
The action of $Q_{-,x}$ on configurations which contain a particle on site $x$ is always zero. Moreover, the $\pm$ sign corresponds to the fermionic string, i.e. it is given by minus one to the number of particles located to the left of $x$.

The requirement that $Q_{-}^2=0$ can be analysed locally by acting with the supercharge twice on the simple configurations shown in \eqref{eqn:localaction}. A detailed analysis shows the only non-vanishing terms come from the first configuration. Imposing that this term vanishes results in the constraint
\begin{equation*}
  a_x c_{x+3} + b_x c_{x}=0.
\end{equation*}
This is a clear analogue of equation \eqref{eqn:lambdaconstraint}, which constrains the parameters in the definition of the supercharge $Q_+$. After having imposed this first restriction, we would like to establish $Q_-$ as a symmetry of our model. Hence we look for values of the parameters $a_x,b_x,c_x$ such that locally it generates the same Hamiltonian as $Q_+$ (i.e. the hopping amplitudes and rules for the potential energy are the same):
\begin{equation} 
  H = \{Q_-,\bar Q_-\}
  \label{eqn:HQminus}
\end{equation}
where $\bar Q_- = Q_-^\dagger$. The explicit comparison of the two sides is a cumbersome task. It leads to a system of quadratic difference equations for the parameters $a_x,b_x,c_x$ which allow to express them in terms of $\lambda_x,\, \mu_x$. Out of the many equations, let us just write
\begin{equation}
  a_{x+2}^2= a_x^2, \quad b_{x+2}^2=b_x^2, \quad c_{x}^2=a_x^2+b_{x+1}^2,
  \label{eqn:constraintsabc}
\end{equation}
which show that once more a staggering with period $2$ is obtained. We skip the details, and report only the solution to the complete set of difference equations which is given by
\begin{equation}
  a_x=  \lambda_{2,2,x} = \lambda_x \mu_{x-1}, \quad b_x = -\lambda_{2,1,x-1} = -\lambda_{x-1}\mu_{x-1}, \quad c_x = \lambda_x.
  \label{eqn:valuesabc}
\end{equation}
Its insertion into \eqref{eqn:constraintsabc} implies that for real positive $\lambda_x, \mu_x$ the parameters of the model lie on the line \eqref{eqn:integrableline}

\paragraph{Translation invariance.} So far we ignored the boundary terms $Q_{-,x}$ with $x=-3,\dots,0$, because all the constraints on the parameters in the definition of $Q_-$ could be derived from local considerations in the bulk. We will now consider the boundary terms and find that it leads to a restriction on the Hilbert space. That is, we find that the supersymmetry $Q_-$ generates the Hamiltonian only in a subsector of the Hilbert space. This feature is quite different from the supersymmetry generated by $Q_+$ which exists on the entire Hilbert space.

To see this we first define the boundary terms by taking advantage of the periodicity of the staggering parameters. To this end, we use the translation operator $T$. Clearly, the staggering with period 2 implies that
\begin{equation}
  T^2Q_{-,x}T^{-2} = Q_{-,x+2}, \quad x=1,\dots,N-2,
  \label{eqn:trslQminus}
\end{equation}
i.e. for the local operators in the bulk. Note that the translation operator on the left (right) of $Q_{-,x}$ acts on states of a chain of length $N+4$ ($N$). We now use this equation to \textit{define} the local operators, $Q_{-,x}$, for $x=-3,\dots,0$ and thus extend \eqref{eqn:trslQminus} to $x=-3,\dots,0$. This definition now also implies that for $x=-3,\dots,0$ we have $Q_{-,x+N} = T^N Q_{-,x}T^{-N}$. Using $T^{-N}=1$ when it acts on a chain of length $N$ and similarly $T^N=T^{-4}$ for a chain of length $N+4$, we can rewrite this as 
\begin{equation*}
  T^4 Q_{-,N+x}= Q_{-,x}, \quad x=-3,\dots,0.
\end{equation*}
From this relation we conclude, in particular, that $Q_-$ is a well-defined mapping between eigenspaces of $T^4$ only when $T^4\equiv 1$ (one easily checks this by computing $T^4 Q_-  T^{-4} $, using the relations above and imposing it to be equal to $Q_-$). Since the Hamiltonian $H = \{Q_+, \bar Q_+\}$ of our model commutes with $T^4$, we conclude that $H = \{Q_-, \bar Q_-\}$ can only hold on subspaces where the translation operator by four sites acts like the identity. Finally, let us mention that as for the non-dynamic supercharge $Q_+$ one may show that the construction of $Q_-$ only works for non-zero twist angles.

\paragraph{Extended supersymmetry algebra.} As we have two copies of the $\mathcal N=2$ supersymmetry algebra in the case of periodic boundary conditions, it appears natural to find two distinct fermion numbers in order to characterise the system. We define the following two operators in terms of the system size $N$ and the original fermion number operator $F$:
\begin{equation*}
  F_V = N-F, \quad \text{and} \quad F_A = F - N/2.
\end{equation*}
Using the mapping to the spin chain, $F_V$ and $F_A$ correspond to the length of the spin chain, and its magnetisation, respectively. With these two fermion numbers, we obtain a lattice representation of the $\mathcal N=(2,2)$ supersymmetry algebra \cite{hori:03} (up to a sign convention for $F_V$) with vanishing central charges
\begin{align}
  & Q_\pm^2=\bar Q_{\pm}^2=0,\quad \{Q_\pm, Q_{\mp}\} = \{Q_\pm,\bar Q_{\mp}\}=0 \nonumber,\\
  & \{Q_\pm, \bar Q_\pm\} = H \pm P   \label{eqn:superalgebra},\\
  & [F_V,Q_\pm]=Q_\pm,\quad  [F_V,\bar Q_{\pm}] = - \bar Q_{\pm} \nonumber,\\
  & [F_A,Q_\pm]=\mp Q_\pm,\quad [F_A,\bar Q_{\pm}] = \pm \bar Q_{\pm}\nonumber,
\end{align}
and zero momentum
\begin{equation*}
  P= 0.
\end{equation*}
For these commutators and anticommutators, the action on chains of appropriate length is implied. Moreover, the algebra exists of course only in translation subsectors with $T^4=1$, where the presence of the dynamic supersymmetry is guaranteed. The fact that the momentum $P$ is zero is consistent with this restriction as we shall see in section \ref{sec:gap} where the relation between the lattice model and its field-theory limit is discussed. 

\subsubsection{Dynamic supersymmetry (2)}
\label{sec:dynneutralsusy}
The dynamic supercharge defined in the last section increases the length of the chain by four sites. A priori, the periodicity of the staggering parameters does not exclude the existence of another length-changing operator which adds only two sites to the system. We will show here that such an operator $Q_0$ exists indeed in certain subsectors of the Hilbert space. The existence of such a symmetry is somewhat expected. In \cite{hagendorf:13} it was shown that the Fateev-Zamolodchikov chain possesses a dynamic supersymmetry of the same structure as the one constructed here below.

\paragraph{Supercharge and Hamiltonian.} Let us show how the operator $Q_0: \mathcal{H}_{N,f} \to \mathcal{H}_{N+2,f+1}$, which inserts two sites and one particle, is constructed. As for the dynamic supercharge $Q_-$, this operator can be written as a sum over local operators
\begin{equation*}
  Q_0 = \sqrt{\frac{N}{N+2}}\sum_{x=-1}^N(-1)^x Q_{0,x}.
\end{equation*}
Unlike for $Q_-$ there is an additional site-dependent string $(-1)^x$: it implies that in the homogeneous limit $Q_0$ inserts a particle with momentum $\pi$ into the system. We will see later that this picture harmonises well with the Bethe-ansatz interpretation of this supersymmetry.

The $Q_{0,x}$ are fermionic operators which insert two sites, and one particle. Like for $Q_{-,x}$ their action is non-vanishing only if the site $x$ is empty. In this case, the action depends on the occupation of the subsequent sites. Let us illustrate the three different possible scenarios by the action on basis vectors for $x=1,\dots,N$:
\begin{align*}
  & Q_{0,x}|\cdots \underset{x}{\textcolor{gray}{\circ}}\hspace{.08cm}\textcolor{gray}{\circ}  \cdots \rangle  = (\pm)\left(\alpha_x|\cdots  \underset{x}{\textcolor{gray}{\circ}}\bullet \circ \hspace{.08cm}\textcolor{gray}{\circ}   \cdots \rangle +\beta_x|\cdots  \underset{x}{\textcolor{gray}{\circ}} \circ \bullet\hspace{.08cm} \textcolor{gray}{\circ}  \cdots \rangle \right)\\
  & Q_{0,x}|\cdots \underset{x}{\textcolor{gray}{\circ}}\bullet \textcolor{gray}{\circ} \cdots \rangle  = (\pm)\left(\gamma_x|\cdots  \underset{x}{\textcolor{gray}{\circ}}\bullet \bullet \circ \textcolor{gray}{\circ}  \cdots \rangle +\delta_x|\cdots  \underset{x}{\textcolor{gray}{\circ}}\circ \bullet\bullet \textcolor{gray}{\circ}  \cdots \rangle \right)\\
  & Q_{0,x}|\cdots \underset{x}{\textcolor{gray}{\circ}}\bullet\bullet \hspace{.08cm}{\textcolor{gray}{\circ}} \cdots \rangle  = (\pm)\left(\epsilon_x|\cdots  \underset{x}{\textcolor{gray}{\circ}} \bullet \circ \bullet \bullet \hspace{.08cm}\textcolor{gray}{\circ}  \cdots \rangle +\eta_x|\cdots  \underset{x}{\textcolor{gray}{\circ}} \bullet \bullet\circ \bullet \hspace{.08cm}\textcolor{gray}{\circ}  \cdots \rangle \right)
\end{align*}
The sign $\pm$ represents the fermionic string, it is given by $-1$ to the number of fermions located to the left of $x$ for $x>1$, and if $x=1$ it is given by $+1$.

The site-dependent parameters, $\alpha_x,\beta_x,\dots, \eta_x$, are constrained by the requirement $Q_0^2=0$. This can be done locally, by acting twice on the three configurations which we use in order to define the action of $Q_{0,x}$, and imposing that the result vanishes locally up to boundary terms. An explicit calculation shows then that this is possible if and only if all parameters are periodic in $x$ with period $2$. Furthermore, it leads to $\beta_x=-\alpha_{x+1},\,\eta_x= \alpha_{x+1}$, and $\epsilon_x = -\alpha_x$, and to the relation
\begin{equation}
  \alpha_x \delta_x + \alpha_{x+1}\gamma_{x+1}=0,
  \label{eqn:relatparams}
\end{equation}
which can be thought of as an analogue of \eqref{eqn:lambdaconstraint}. The periodicity of the parameters leads to the following relation
\begin{equation*}
 T^2 Q_{0,x}T^{-2} = Q_{0,x+2},\quad x=1,\dots,N-2.
\end{equation*}
The remaining operators $Q_{0,-1},\,Q_{0,0}$ which take into account the insertion of a pair of sites between $x=N$ and $x=1$ are defined by extending this relation to $x=-1,0$. Using a similar reasoning as for the dynamic supercharge $Q_{-}$ presented above, we find that they can be equivalently expressed as
\begin{equation*}
  T^2 Q_{0,x+N} = Q_{0,x}, \quad x=-1,0. 
\end{equation*}
In complete analogy with the case of $Q_-$ we conclude from this equation that the supercharge $Q_0$ is a well-defined mapping between translation sectors only if $T^2\equiv 1$. In these subsectors it generates a supersymmetric Hamiltonian 
\begin{equation}
  H = \{Q_0,\bar Q_0\}.
  \label{eqn:HamiltonianQ0}
\end{equation}
Let us count how many remaining parameters there are at this stage: we are left with $\alpha_x,\gamma_x,\delta_x$ which are periodic under $x\to x+2$, and subject to \eqref{eqn:relatparams}. Taking into account that we are free to rescale them, we find thus three free parameters. Their number is furthermore reduced if we impose that the Hamiltonian $H$ coincide with the one for the $M_2$ model. Quite interestingly, it implies also a restriction to the line of couplings \eqref{eqn:integrableline}, and the $2$-periodicity for $\lambda_x, \mu_x$. Indeed, the analysis of the potential energies on both sides of \eqref{eqn:HamiltonianQ0} shows that equality can hold only if
\begin{equation}
  \gamma_x^2 + \delta_x^2 = 4\alpha_{x+1}^2.
  \label{eqn:constraint}
\end{equation}  
Adjusting the hopping terms one finds that the parameters are related to the original staggering parameters according to
\begin{equation*}
  \alpha_x = \frac{\lambda_{x+1}}{2},\quad \beta_x = -\frac{\lambda_x}{2} ,\quad \gamma_x = \lambda_x \mu_x, \quad \delta_x = \lambda_x \mu_{x+1}, \quad \eta_x = \frac{\lambda_x}{2},\quad \epsilon_x=-\frac{\lambda_{x+1}}{2}.
\end{equation*}
This implies periodicity, and via reinsertion into \eqref{eqn:constraint} the special line \eqref{eqn:integrableline}. Furthermore, we checked explicitly that in the subsectors where $T^2\equiv 1$, the dynamic supercharge $Q_0$ anticommutes with the other ones:
\begin{equation*}
  \{Q_0,Q_\pm\}=0, \quad \{\bar Q_0,Q_\pm\}=0.
\end{equation*}


\section{The coordinate Bethe ansatz}
\label{sec:ba}
In this section, we show that the Hamiltonian of the $M_2$ model can be diagonalised by means of the coordinate Bethe ansatz along a two-parameter submanifold in the space of staggering parameters which coincides precisely with the submanifold with enhanced supersymmetry identified in the previous section.

The technique employed here below is a combination of the Bethe-ansatz for the staggered $M_1$ model found by Nienhuis and Blom \cite{blom:12}, the coordinate Bethe ansatz for higher spin XXX chains \cite{crampe:11}, and an asymptotic analysis which allows an easy determination of the integrable manifold in the space of parameters. We start by specifying the basis, and Bethe-ansatz form for the wave function in section \ref{sec:basis}. In section \ref{sec:oneparticle} we analyse the one-particle problem: it is not directly solvable for general staggering, but will allow to determine some useful asymptotic expansions of the single-particle wave function. The two-particle problem is solved in section \ref{sec:twoparticles}. We show that it fixes the choice of admissible staggering parameters as well as the period of the staggering to the special submanifold. Furthermore, we introduce a useful elliptic parametrisation for it. The many-particle case is considered afterwards, and leads to the Bethe ansatz equations.

\subsection{Basis vectors and Bethe-ansatz form of the wave function}
\label{sec:basis}
\paragraph{Basis vectors.} Our principal goal is to diagonalise the Hamiltonian $H$, i.e. to solve the Schr\"odinger equation
\begin{equation}
  H|\psi\rangle = E|\psi\rangle.
  \label{eqn:schroedinger}
\end{equation}
To this end we need to choose a suitable basis in the fermion Hilbert space which simplifies our problem as much as possible. The most natural choice appears to be 
the canonical occupation number basis: a basis vector is labeled by the positions of the particles in a given configurations:
\begin{equation*}
  |x_1,\dots,x_f\rangle = |\circ \cdots \circ \underset{x_1}{\bullet}\circ\cdots \circ\underset{x_f}{\bullet}\circ \cdots \circ\rangle.
\end{equation*}
This basis is orthonormal which is convenient for many applications. However, the Bethe ansatz is more conveniently formulated when using basis vectors $||x_1,\dots,x_f\rrangle$ which differ from the canonical ones by configuration-dependent factors. Let us denote by $x_1',x_2',\dots$ the positions of the first members of pairs in the configuration $x_1,x_2,\dots, x_f$. We introduce the non-orthonormal basis
\begin{equation}
  ||x_1,x_2,\dots, x_f\rrangle = \left(\prod_{j}C_{x_j'}\right)|x_1,x_2,\dots, x_f\rangle
  \label{eqn:nonorthonormalbasis}
\end{equation}
with normalisation factors $C_{x}$ to be determined. This modified basis is similar to the basis used in the coordinate Bethe ansatz solution for the higher-spin XXX chains studied in \cite{crampe:11}. It allows to absorb a trivial part of the wave function into the basis itself.

\paragraph{Bethe ansatz.} The Hamiltonian of our model, twisted or not, commutes obviously with the fermion number operator $F$, and can therefore be diagonalised separately in each subsector $\mathcal H_{N,f}$. We expand its eigenstates in $\mathcal H_{N,f}$ in the modified basis
\begin{equation}\label{eqn:psi}
  |\psi\rangle = \sum_{\{x\}}\psi(x_1,x_2,\dots, x_f)||x_1,x_2,\dots, x_f\rrangle.
\end{equation}
Here, the sum is taken over all positions $1\leq x_1<x_2<\dots <x_f\leq N$ of the particles which respect the exclusion constraint of the $M_2$ model, and the boundary conditions. For the wave function $\psi(x_1,\dots, x_f)$ we make the Bethe ansatz by writing it as a linear combination of products of single-particle wave functions $\varphi(x;z)$:
\begin{equation}
  \psi(x_1,\dots,x_f) = \sum_{\sigma \in S_f}B_\sigma \varphi(x_1;z_{\sigma(1)})\cdots \varphi(x_f;z_{\sigma(f)}).
  \label{eqn:betheansatz}  
\end{equation}
The sum is over all permutations $\sigma$ of $f$ objects, weighted by certain amplitudes $B_\sigma$. The variables $z_1,\dots, z_f$ are the rapidities of the particles. In order to give meaning to these rapidity variables we need to specify the structure of the single-particle wave functions. In fact, we shall assume that the weights of the model are periodic: $\lambda_{x+p}=\lambda_x,\, \mu_{x+p}=\mu_x$. Hence, single particles are described by Bloch wave functions
\begin{equation}
  \varphi(x;z) = A_x(z)z^x, \quad A_{x+p}(z)=A_x(z).
  \label{eqn:BlochWF}
\end{equation}
Our main objective here is to show that the model is Bethe-ansatz solvable if and only if $p=2$, and the staggering parameters are chosen from the submanifold \eqref{eqn:integrableline}.

Given the basis and Bethe ansatz form of the wave function we proceed now through a series of standard steps for the solution of the Schr\"odinger equation \eqref{eqn:schroedinger}. We project this equation on simple basis vectors, and resolve the resulting system of difference equations for the wave functions $\psi(x_1,\dots, x_f)$. In the following sections, we address first the case of $f=1$ and $2$ particles, mention briefly the case of $f=3$ and $4$ particles, and deduce the result for general $f$ through a standard argument.

\subsection{The one-particle problem}
\label{sec:oneparticle}

In the subsector of the Hilbert space where no particles are present the diagonalisation of the Hamiltonian is trivial: the empty state $|\circ \circ \cdots \circ\rangle$ is an eigenvector of $H$ with eigenvalue $E=\sum_{x=1}^N\lambda_x^2$. It serves as a reference state for the Bethe ansatz and will allow to build eigenstates with non-zero fermion numbers. Let us start with a single particle $f=1$. We write thus the eigenvalue as
\begin{equation*}
  E = \sum_{x=1}^N\lambda_x^2 + \epsilon(z),
\end{equation*}
where $\epsilon(z)$ denotes the excitation energy for a (pseudo-)particle with rapidity $z$ above the reference-state level. The corresponding Bethe-ansatz wave function is in fact simply given by $\varphi(x;z)$. Using its Bloch-wave structure \eqref{eqn:BlochWF} we find the difference equation
\begin{align}
  (\epsilon(z) + \lambda_{x-1}^2(1-\mu_{x-1}^2) + &\lambda_{x+1}^2(1-\mu_{x}^2))A_x(z)\label{eqn:recursionA}\\
   &= \lambda_x\lambda_{x+1}(1-\mu_x^2) A_{x+1}(z)z + \lambda_x\lambda_{x-1}(1-\mu_{x-1}^2) A_{x-1}(z)z^{-1}\nonumber
\end{align}
Given the periodicity in $x$, we conclude that this leads to a system of $p$ homogeneous linear equations for the quantities $A_1(z),\dots, A_p(z)$. In order to have a non-trivial solution its coefficient matrix needs to have zero determinant. This leads to a polynomial equation of order $p$ for the excitation energy, and determines the dispersion relation, $\epsilon=\epsilon(z)$. Without knowing $p$ it is not very useful to write down this system and its solution explicitly. Instead, we will analyse it in the formal limit where $z\to 0$, and $z\to \infty$, in order to obtain an asymptotic expansion for $\epsilon(z)$ and the amplitude ratio
\begin{equation*}
  f_x(z) = \frac{\lambda_x A_{x+1}(z)z}{\lambda_{x+1}A_x(z)}.
\end{equation*}
Let us start with large rapidity. It is clear from \eqref{eqn:recursionA} that both the excitation energy, and the amplitude ratio diverge linearly for large $z$. Writing
\begin{equation*}
  \epsilon(z) = \gamma_{-1} z + \gamma_0 + O(z^{-1}), \quad f_x(z) = \alpha_x z (1+ \beta_{x} z^{-1} + O(z^{-2}))
\end{equation*}
we find that the coefficients $\alpha_x,\beta_x$ are given by
\begin{equation*}
  \alpha_x = \frac{\gamma_{-1}}{\lambda_{x+1}^2(1-\mu_x^2)}, \quad \beta_x = \frac{\gamma_0 + \lambda_{x+1}^2(1-\mu_x^2)+\lambda_{x-1}^2(1-\mu_{x-1}^2)}{\gamma_{-1}}.
\end{equation*}
Here the constants $\gamma_{-1}$ and $\gamma_0$ may in principle be determined from the periodicity $A_{x+p}(z)=A_{x}(z)$ which leads to $f_1(z)f_2(z)\cdots f_p(z)=1$, but we will not need their explicit form. The limit of small rapidity leads to similar results. We find that the excitation energy and amplitude ratio have the expansions
\begin{equation*}
  \epsilon(z) = \delta_{-1} z^{-1} + \delta_0 + O(z), \quad f_x(z) = \rho_x z (1+ \eta_{x} z + O(z^2))
\end{equation*}
where the coefficients $\rho_x,\beta_x$ are given by
\begin{equation*}
  \rho_x = \frac{\lambda_{x}^2(1-\mu_x^2)}{\delta_{-1}}, \quad \eta_x = -\frac{\delta_0 + \lambda_{x}^2(1-\mu_x^2)+\lambda_{x+2}^2(1-\mu_{x+1}^2)}{\gamma_{-1}}.
\end{equation*}

It would be well justified to question the use of these expansions at this point. The idea is the following. For the two-particle problem we will derive the $S$-matrix of the model as a complicated combination of the amplitude ratios $f_x(z_{j}),\,j=1,2$. The expression carries an $x$-dependence which should however be spurious. The formal limit where one of the rapidities tends to zero or infinity allows to understand the (pole) structure of the $S$-matrix, and determine conditions on the staggering parameters which yield a position-independent $S$-matrix. The expressions derived here above will be instrumental in this procedure.

\subsection{The two-particle problem and the S-matrix}
\label{sec:twoparticles}

Next, we consider the case of two particles $f=2$. As long as we project the Schr\"odinger equation on configurations where the two particles are far apart (and far from the boundaries), the Bethe-ansatz wave function \eqref{eqn:betheansatz} solves the resulting difference equation with the eigenvalue
\begin{equation}
  E= \sum_{x=1}^N \lambda_x^2 + \epsilon(z_1) + \epsilon(z_2).
  \label{eqn:E2particles}
\end{equation}
If, however, the particles are next-to-nearest or nearest neighbours, we have to take into account that the action of the Hamiltonian induces pair formation, pair splitting, and pair hopping.

\subsubsection{Next-to-nearest neighbours}
We start with the projection of the Schr\"odinger equation on a configuration where two particles are next-to-nearest neighbours $\cdots \circ \bullet \circ \bullet \circ \cdots$, i.e. on some basis vector $|x,x+2\rangle$. For this case we find the rather long equation
\begin{align*}
  E\psi(x,x+2) =& \,\lambda_x\lambda_{x-1}(1-\mu_{x-1}^2)\psi(x-1,x+2) + \lambda_{x+2}\lambda_{x+3}(1-\mu_{x+2}^2)\psi(x,x+3)\\
  &+ \lambda_x\lambda_{x+1}\mu_{x+1}C_{x+1}\psi(x+1,x+2)+ \lambda_{x+1}\lambda_{x+2}\mu_{x}C_{x}\psi(x,x+1)\\
  &+\left(\sum_{y=1}^N \lambda_y^2 -\lambda_{x-1}^2(1-\mu_{x-1}^2)-\lambda_{x+1}^2 - \lambda_{x+3}^2(1-\mu_{x+2}^2)\right)\psi(x,x+2).
\end{align*}
If the Bethe ansatz holds together with the form of the eigenvalue as written in \eqref{eqn:E2particles} then each of the arguments of the wave function can formally be treated as for isolated particles. This leads to a second eigenvalue equation. Equating the two expressions we find the difference equation
\begin{align*}
  \lambda_x\lambda_{x+1}(\mu_{x+1}C_{x+1}+\mu_x^2-1)&\psi(x+1,x+2)\\
&+\lambda_{x+1}\lambda_{x+2}(\mu_{x}C_{x}+\mu_{x+1}^2-1)\psi(x+1,x+2)\\
  &+\lambda_{x+1}^2(1-\mu_x^2-\mu_{x+1}^2)\psi(x,x+2)=0.
\end{align*}
Given this equation there are two ways to proceed. We could impose the equation as a constraint on the wave function, leaving the normalisation factors $C_x$ undetermined, but fixing the structure of the $S$-matrix. However, one can show that this leads to a contradiction for other particle arrangements. Therefore, we will instead require that the coefficients multiplying the wave functions in this equation be identically zero, so that the equation does not lead to any constraints for the $S$-matrix \cite{crampe:11}. In the present case, the only non-trivial solution is \begin{equation}
  C_x = \mu_x, \quad \text{and} \quad \mu_{x}^2+\mu_{x+1}^2=1,
  \label{eqn:cxandrelationmu}
\end{equation}
for arbitrary $x$. We conclude that our requirement implies the $2$-periodicity of the staggering parameters $\mu_x$ while it does not fix the $\lambda_x$. As we shall see, the latter will be constrained by the nearest-neighbour problem.

\subsubsection{Nearest neighbours}
Next, we project the Schr\"odinger equation for two particles onto a configuration with a single pair $\cdots \circ \circ \bullet \bullet \circ \circ \cdots$, i.e. on the basis vector $|x,x+1\rangle$.
We follow the same procedure as in the last section, and find the following difference equation for the wave function
\begin{align}
  &\lambda_{x-1}(\lambda_{x+1}\psi(x-1,x)-\lambda_x\psi(x-1,x+1)+\lambda_{x-1}\psi(x,x+1))+\lambda_x\lambda_{x+1}\psi(x,x) \nonumber\\
  &+ \lambda_{x+2}(\lambda_{x+2}\psi(x,x+1)-\lambda_{x+1}\psi(x,x+2)+\lambda_{x}\psi(x+1,x+2))\nonumber\\
&+\lambda_x\lambda_{x+1}\psi(x+1,x+1)=0.\label{eqn:diffequWF}
\end{align}
Unlike in the case of next-to-nearest neighbours this equation cannot vanish identically for non-trivial choices of the staggering parameters. Hence it will lead to constraints on the parameters in the Bethe ansatz wave function. Indeed, using \eqref{eqn:betheansatz} we find that the two amplitudes $B_{12}$ and $B_{21}$ are related by
\begin{equation}
  B_{12} P_x(z,w) + B_{21} P_x(w,z)=0,
  \label{eqn:amplitudeequation}
\end{equation}
where $P_x(z,w)$ denotes the complicated expression
\begin{align*}
  P_x(z,w) = \lambda_{x-1}^2\left(\frac{1-f_x(w)}{f_{x-1}(z)}+f_x(w)\right)&+ \lambda_x^2+ \lambda_{x+1}^2 f_x(z)f_x(w)\\
  & + \lambda_{x+2}^2\left(f_x(w)f_{x+1}(w)(f_x(z)-1)+f_x(w)\right).
\end{align*}
Let us suppose that $z$ and $w$ are such that both $P_x(z,w)$ and $P_x(w,z)$ are non-vanishing for all $x$. In this case we find the $S$-matrix of the model
\begin{equation*}
  S(z,w) = \frac{B_{12}}{B_{21}} = - \frac{P_x(w,z)}{P_x(z,w)}.
\end{equation*}
The fact that this expression is rather implicit, since we have not yet found a general expression for $f_x(z)$, is perhaps less dramatic than its $x$-dependence. Indeed, the Bethe ansatz assumes that the amplitudes $B_\sigma$ are position-independent. For generic choices of the staggering parameters the formula for $S(z,w)$ leads, however, to an $x$-dependent expression. It follows that the only possible choice for the model to be Bethe-ansatz solvable is to \textit{impose} the site-independence for arbitrary $z,w$. One may try to do this by working directly with the given expression, but the resulting equations are quite involved. Hence, we choose to take advantage of our asymptotic expansions for $f_x(z)$ derived in section \ref{sec:oneparticle}, and analyse the limit $z\to \infty$. We find
\begin{equation*}
  S(z,w) = z \sigma(w) + O(1), \quad \text{with} \quad \sigma(w) = \frac{\gamma_{-1}(1-f_x(w))}{\mu_x^2f_x(w)(\lambda_{x+1}^2+\lambda_{x+2}^2f_{x+1}(w))}.
\end{equation*}
Like $S(z,w)$ the function $\sigma(w)$ has to be position-independent. Finding which staggering parameters lead to this independence is still a delicate task, and hence we analyse again only the leading terms of the expansion of $\sigma(w)$ as $w \to 0$:
\begin{equation*}
  \sigma(w) =\frac{\delta_{-1}}{\mu_x^2\mu_{x+1}^2 \lambda_x^2\lambda_{x+1}^2}\left(\delta_{-1}w^{-1}-\delta_0+O(w)\right).
\end{equation*}
Every term in this expansion needs to be independent of $x$. Using the $2$-periodicity for $\mu_x$, we see that this can at leading order only be true if the product $\lambda_x^2\lambda_{x+1}^2$ is independent of $x$. This condition implies trivially that for real positive staggering parameters we have
\begin{equation*}
  \lambda_{x+2} = \lambda_x.
\end{equation*}
Hence, we find that the expression for the $S$-matrix found above is independent of the position $x$ only if the staggering parameters are $2$-periodic, and satisfy \eqref{eqn:integrableline}, which confirms that the latter is a necessary condition for Bethe-ansatz solvability.

\paragraph{Reduced difference equation.} If all staggering parameters have period two, then the difference equation for the wave function can be simplified. Indeed, notice that for $\lambda_{x+2}=\lambda_x$ the first four terms equal the last four terms of the left-hand side of \eqref{eqn:diffequWF} up to a shift $x\to x+1$. We may formalise this by introducing the shift operator $\mathcal T$ defined through $\mathcal Tf(x)=f(x+1)$. Then we find
\begin{equation}
  (1+\mathcal T)\left[\lambda_{x-1}(\lambda_{x+1}\psi(x-1,x)-\lambda_x\psi(x-1,x+1)+\lambda_{x-1}\psi(x,x+1))+\lambda_x\lambda_{x+1}\psi(x,x) \right]=0\nonumber 
\end{equation}
Hence the expression within brackets lies in the kernel of $(1+\mathcal T)$, i.e. it is of the form $(-1)^x \times \text{const.}$ In fact, one can show that this constant needs to be zero (we omit this tedious and not very illuminating discussion here). Using this result we find the reduced difference equation
\begin{equation}
  \lambda_{x+1}(\psi(x-1,x)+\psi(x,x+1))+\lambda_x(\psi(x,x)-\psi(x-1,x+1)) =0.
  \label{eqn:reducedrecursion}
\end{equation}
This relation has two advantages. First of all, it allows to write a somewhat simpler version of \eqref{eqn:amplitudeequation}. We find that
\begin{subequations}
\begin{equation}
  B_{12}R_x(z,w) + B_{21}R_x(w,z)=0
\end{equation}
where $R_x(z,w)$ is given by the expression
\begin{equation}
  R_x(z,w)= 1+\frac{\lambda_{x+1}^2}{\lambda_x^2}\left(\frac{1-f_x(w)}{f_{x-1}(z)}+f_x(w)\right).
\end{equation}
\label{eqn:simplifiedamplitudeequ}%
\end{subequations}
This leads to a simplified expression for the $S$-matrix, and will be used below to derive a closed expression for it in terms of Jacobi theta functions.
Second, \eqref{eqn:reducedrecursion} proves to be quite useful in order to show that processes involving three and four particles are indeed coherent.

\subsubsection{Elliptic parametrisation}
In the previous section, we saw that it is necessary to restrict the parameters of the model to the special submanifold. The aim of this and the following section is to show that this restriction is also sufficient for the model to be Bethe-ansatz solvable. To this end, it is convenient to re-examine the one-particle problem, and determine an explicit parametrisation for the rapidities and the excitation energy, which will lead to an explicit and simple form for the $S$-matrix.

\paragraph{Staggering parameters.} We need a suitable parametrisation of the parameters $\lambda_x$ and $\mu_x$. It turns out that a convenient choice is to write them in terms of Jacobi theta functions $\vartheta_j(u) = \vartheta_j(u,q),\, j=1,\dots,4$ where $q$ is the so-called elliptic nome. We follow the conventions of Whittaker and Watson \cite{whittaker:27}. In fact, it is sufficient to define
\begin{equation*}
  \vartheta_1(u,q) = -\i \sum_{j=-\infty}^\infty (-1)^j q^{(j+1/2)^2}e^{(2j+1)\i u}.
\end{equation*}
The other theta functions are obtained by shifting the argument by $\pi/2, \pi \tau/2,\pi/2+\pi \tau/2$ where $\tau$ is related to the elliptic nome according to $q= e^{\i \pi \tau}$. For instance, $\vartheta_4(u) = \i q^{1/4} e^{-\i u}\vartheta_1(u-\pi\tau/2)$. These functions can be thought of as generalisations of the trigonometric or the exponential functions. They satisfy a host of identities, in particular various addition theorems which are at the heart of the simplifications in the following.

\begin{figure}[h]
\centering
\includegraphics[width=0.65\columnwidth]{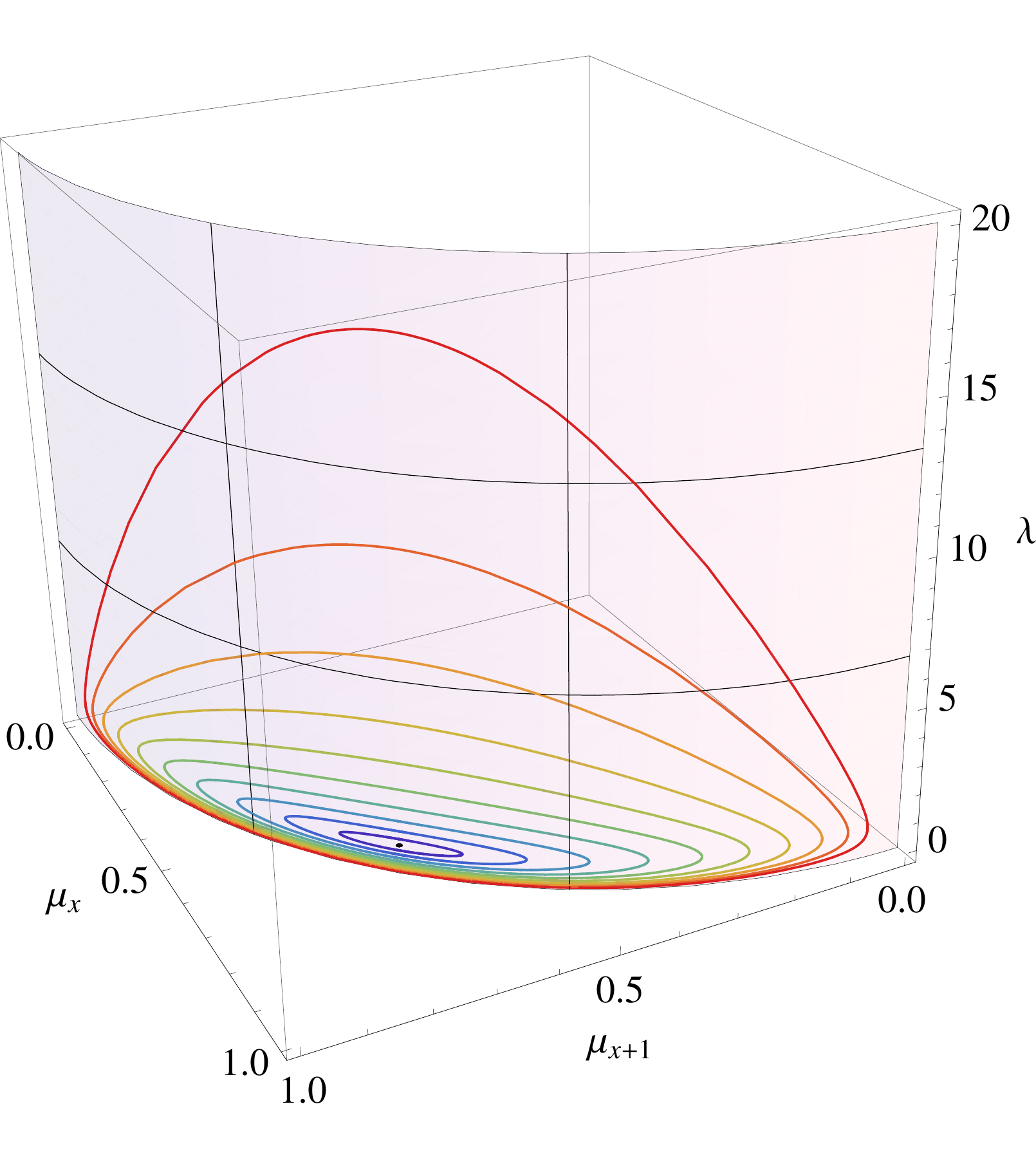}
\caption{We plot the parameters $\mu_x, \mu_{x+1}$ and $\lambda=\lambda_x/\lambda_{x+1}$ as a function of $t$ with $0\leq t\leq 2 \theta$ for $q=0, 0.05, \dots,0.5$. The black dot corresponds to $q=0$, the outer red line corresponds to $q=0.5$. Finally, we also show the special submanifold given by \eqref{eqn:integrableline}, where the model is integrable and enjoys additional supersymmetry. The constant $q$ lines lie on the special submanifold.}
\label{fig:tparam}
\end{figure}

The staggering parameters are given as functions of two real parameters: $t$ and the elliptic nome $0\leq q<1$. In terms of theta functions they read
\begin{equation*}
  \mu_x^2 = \left(\frac{\vartheta_1(\theta)}{\vartheta_1(2\theta)}\right)^2\frac{\vartheta_4(t+2x\theta)^2}{\vartheta_4(t+(2x-1)\theta)\vartheta_4(t+(2x+1)\theta)}, \quad \theta = \frac{\pi}{4},
\end{equation*}
and
\begin{equation*}
  \lambda_x^2= 2\left(\frac{\vartheta_1(\theta)}{\vartheta_1(2\theta)}\right)^2\frac{\vartheta_4(t+(2x-1)\theta)^2}{\vartheta_4(t+2(x-1)\theta)\vartheta_4(t+2x\theta)}, \quad \theta = \frac{\pi}{4}.
\end{equation*}
One may check that this choice is compatible with both periodicity, and the equation $\mu_x^2+\mu_{x+1}^2=1$. The second equation implies that $\lambda_x^2+\lambda_{x+1}^2=2$, a normalisation which we are free to choose. The latter is designed to recover $\lambda_x=1$ and $\mu_x=1/\sqrt{2}$ in the trigonometric limit where the elliptic module $q$ tends to zero, and the Hamiltonian becomes translation invariant. For non-zero $q$, we note that the transformation $t \to t+ 2\theta$ is equivalent to a shift $x \to x+1$, and thus a translation of the system by one site. The elliptic parametrisation given here can be justified and derived in a systematic analysis of the $M_k$ models for all $k=1,2,3,\dots$ \cite{chlh:tbp}. In figure \ref{fig:tparam} we plot the parameters as a function of $t$ with $0\leq t\leq 2 \theta$ for $q=0, 0.05, \dots,0.5$. It is clear that this parametrisation maps out the special submanifold.

\paragraph{Excitation energy, rapidity and one-particle wave function.} Now let us use this parametrisation in order to find convenient expressions for the excitation energy $\epsilon(z)$. As we know, it is determined by the homogeneous linear $p\times p$ system \eqref{eqn:recursionA}. The period $p=2$ of the staggering parameters implies that the periodic part of the single-particle Bloch wave function satisfies $A_{x+2}(z)=A_x(z)$. We may use this in the recursion relation \eqref{eqn:recursionA} which becomes of first order (as opposed to second order for arbitrary $p>2$):
\begin{align*}
  (\epsilon(z) +  \lambda_{x+1}^2)A_x(z)=\lambda_x\lambda_{x+1}\left(\mu_{x+1}^2 z + \mu_x^2 z^{-1}\right)A_{x+1}(z)
\end{align*}
Applying the periodicity property after shifting $x\to x+1$, we obtain that $\epsilon(z)$ solves the second-order polynomial equation
\begin{equation*}
  \epsilon(z)(\epsilon(z)+2) = \Lambda^2(z-z^{-1})^2, \quad \Lambda = 2\left(\frac{\vartheta_1(\theta)}{\vartheta_1(2\theta)}\right)^4.
\end{equation*}
Notice that this equation is independent of the parameter $t$. As we shall see the spectrum of the Hamiltonian is $t$-independent as a consequence. Solving this equation for $\epsilon(z)$ leads to roots of quartic polynomials in $z$ which are neither elegant nor useful. Instead, we will uniformise this equation through the introduction of theta-function parametrisations of the rapidities. One checks that the choice 
\begin{equation}
  z(u) = - \frac{\vartheta_1(u+\theta)}{\vartheta_1(u-\theta)}
  \label{eqn:rapidityparam}
\end{equation}
leads to
\begin{equation}
  \epsilon(u) \equiv \epsilon(z(u)) = -2\left(\frac{\vartheta_1(\theta)}{\vartheta_1(2\theta)}\right)^2\frac{\vartheta_1(u)^2}{\vartheta_1(u-\theta)\vartheta_1(u+\theta)}.
  \label{eqn:excitationenergy}
\end{equation}
The main tools in all these calculations are the addition theorems for the Jacobi theta functions mentioned above.
Using these two relations, one may determine the functions $A_x(z)$ as a function of the parameter $u$ from the recursion relation given above up to an overall factor. We fix the latter by the requirement $A_x(z=1) = \lambda_x$. This gives
\begin{equation*}
  A_x(u) \equiv A_x(z(u)) = \frac{\lambda_x \vartheta_4(t-u +(2x-1)\theta)}{\vartheta_4(t+(2x-1)\theta)}.
\end{equation*}

\paragraph{The S-matrix.}
We are now in the position to derive an explicit expression for the $S$-matrix in terms of Jacobi theta functions. To this end, we use \eqref{eqn:simplifiedamplitudeequ}, and express all the amplitude ratios $f_x(z)$ in terms of the expressions given in the last paragraph. We find that
\begin{equation*}
  R_x(z(u),z(v))=-\frac{\vartheta_1(2\theta)^2\vartheta_4(t+2x\theta)\vartheta_4(t+2(x{-}1)\theta)\vartheta_4(t-(u{+}v)+(2x{-}1)\theta)}{\vartheta_1(\theta)\vartheta_4(t+(2x{-}1)\theta)\vartheta_4(t-u+(2x{-}1)\theta)\vartheta_4(t-v+(2x{-}1)\theta)} r(u,v)
\end{equation*}
where $r(u,v)$ has the simple form
\begin{equation}
  r(u,v) = \frac{\vartheta_1(u-v+\theta)}{\vartheta_1(u+\theta)\vartheta_1(v-\theta)}.
  \label{eqn:rofuandv}
\end{equation}
Notice that the $x$-dependent part is symmetric in $u,v$, and that $r(u,v)$ does not depend on the parameter $t$. This implies that we are left with the rather simple equation $B_{12} r(u,v) + B_{21} r(v,u)=0$, and hence find the $S$-matrix
\begin{equation*}
  S(u,v) \equiv S(z(u),z(v)) = \frac{z(u)}{z(v)}\frac{\vartheta_1(u-v-\theta)}{\vartheta_1(u-v+\theta)}, \quad \theta = \frac{\pi}{4}.
\end{equation*}
This expression does not have the difference property, i.e. it does not depend on $u,v$ only through the difference $u-v$. The reason for this is the exclusion constraint of the model, which leads to the prefactor $z(u)/z(v)$. Otherwise, the expression is similar to the one for the eight-vertex model. The result of Blom and Nienhuis for the $M_1$ model is similar \cite{blom:12}, one needs to choose $\theta = \pi/3$ instead of $\theta=\pi/4$.

\subsection{Many particles: boundary conditions and the Bethe equations}
\label{sec:manyparticles}

\paragraph{Consistency.} Considering the one- and two-particle problems is not sufficient to conclude that the model is Bethe-ansatz solvable. The next level of difficulty comes from testing the Bethe-ansatz for local three- and four-particle interactions, i.e. for projections on configurations $\cdots \circ \bullet \bullet \circ \bullet \circ \cdots$ and $\cdots \circ \bullet \circ \bullet \bullet \circ \cdots$ in the sector with $f=3$ particles, and $\cdots \circ \bullet \bullet \circ \bullet \bullet \circ \cdots$ in the sector with $f=4$ particles.

The key to show consistency for these situations is the reduced difference equation \eqref{eqn:reducedrecursion}, which we derived for the two-particle wave function. In fact, the nature of the Bethe ansatz implies that it holds in fact for any $f$ in the following way
\begin{align}
  \lambda_{x+1}(\psi(\dots, x-1,& x,\dots)+ \psi(\dots,x,x+1,\dots))\nonumber\\
  &+ \lambda_{x}(\psi(\dots, x,x,\dots)- \psi(\dots,x-1,x+1,\dots))=0.
  \label{eqn:rec}
\end{align}

Let us start with $f=3$ particles. If we consider \eqref{eqn:rec} with an additional particle on the site $x+1$ or $x+2$, and re-apply it to the existing two particles, we obtain quite straightforwardly the following equations for the three-particle wave function:
\begin{equation}
  \psi(x-1,x+1,x+1)+\psi(x,x+1,x+2)=\psi(x-1,x,x+1)+\psi(x,x,x+2)=0.
  \label{eqn:strangerec}
\end{equation}
These equations are sufficient to prove that the three-particle problem is consistent. Indeed, when projecting the Schr\"odinger equation onto the basis vector $|x,x+2,x+3\rangle$, and comparing it as for the one- and two-particle problems, to the case when all particles are treated as if they were free, one obtains the following constraint on the wave function:
\begin{align*}
 \lambda_{x+1}(\psi(x,&x+2,x+4)-\psi(x, x+2,x+2))\\
&-\lambda_x(\psi(x,x+3,x+4)+\psi(x,x+2,x+3))\\
  &\quad \quad =\lambda_x(\psi(x+1,x+1,x+3)+\psi(x+1,x+2,x+3)).
\end{align*}
The left-hand side vanishes as a consequence of \eqref{eqn:rec} whereas the vanishing of the right-hand side is due to \eqref{eqn:strangerec}. The projection on the basis vector $|x,x+1,x+3\rangle$ leads to the same type of relation which holds identically. This exhausts all possible three-particle interactions, and shows that they can be reduced to two-particle processes via \eqref{eqn:rec}.

The case to be checked for $f=4$ particles is the projection of the Schr\"odinger equation on the state $|x,x+1,x+3,x+4\rangle$. The comparison to the case of free particles leads to the consistency constraint
\begin{align*}
  &\mu_{x+1}^2\Bigl(\lambda_x(\psi(x-1,x+1,x+3,x+4)-\psi(x,x,x+3,x+4))\\
  &\hspace{3cm}-\lambda_{x+1}(\psi(x-1,x,x+3,x+4)+\psi(x,x+1,x+3,x+4))\Bigr)\\
  &+\mu_{x}^2\Bigl(\lambda_x(\psi(x,x+1,x+3,x+5)-\psi(x,x+1,x+4,x+4))\\
  &\hspace{3cm}-\lambda_{x+1}(\psi(x,x+1,x+4,x+5)+\psi(x,x+1,x+3,x+4))\Bigr)\\
  &=\lambda_x\Bigl(\mu_{x+1}^2(\psi(x+1,x+1,x+3,x+4)+\psi(x,x+1,x+2,x+4)\\
  &\hspace{3cm}+\mu_{x}^2(\psi(x,x+2,x+3,x+4)+\psi(x,x+1,x+3,x+3)\Bigr).
\end{align*}
The left-hand side of this lengthy equation vanishes by applying \eqref{eqn:rec} to the second two variables, whereas the right-hand side gives zero by application of \eqref{eqn:strangerec} to the first and last three variables. Hence, also the four-particle problem is consistent. This exhausts all cases which need to be checked: the consistency for configurations with higher particle numbers can be reduced to linear superpositions the one-, two-, three- and four-particle situations. We conclude that the Bethe-ansatz works consistently for the staggered $M_2$ model along the special line in parameter space \eqref{eqn:integrableline}.

\paragraph{Translation symmetry.} 
We consider from now on an arbitrary number of fermions $f>1$  on the chain. In order to discuss the Bethe-ansatz equations we need to specify the boundary conditions of our model. We will consider the model with a twist. Because of the $2$-periodicity of the staggering, the Hamiltonian commutes with the square of the twisted translation operator $T'$, $[H,(T')^2]=0$, as explained in section \ref{sec:defmodel}. Hence we impose the solutions of the Schr\"odinger equation to be eigenvectors of $(T')^2$:
\begin{equation*}
  (T')^2|\psi\rangle = \mathfrak{t}^2|\psi\rangle.
\end{equation*}
The equation is written in a suggestive form: for the homogeneous model $\mathfrak t$ is simply the eigenvalue of $T'$, which becomes a proper symmetry of the Hamiltonian in this case.
In the general, staggered case, we use \eqref{eqn:psi} and project the resulting equation on a configuration $|x_1,\dots,x_f\rangle$. Assuming that the first $m=0,1,2$ particles are located on the first 2 lattice sites, the projection leads to
\begin{align*}
  (-1)^{m(f-1)}e^{\i \phi(m-2f/N)}\psi(x_{m+1}-2,\dots, x_f-2&,x_1+N-2,\dots, x_m+N-2)\\
     &= \mathfrak t^2 \psi(x_1,\dots, x_f).
\end{align*}
In the case $m=0$, we use the Bethe ansatz \eqref{eqn:betheansatz} and conclude that the eigenvalue $\mathfrak t^2$ is given by
\begin{equation}
  \mathfrak t^2 = e^{-2i\phi f/N}\prod_{j=1}^f z_j^{-2}.
  \label{eqn:eigenvaluetransop}
\end{equation}
If we require this result to be compatible with the other choices $m=1,2$ then we obtain a common constraint on the transformation behaviour of the amplitudes $B_\sigma$ under a cyclic shift of the permutation $\sigma$. Let $\pi$ be the cyclic shift, i.e. $\pi(1)=2,\pi(2)=3,\dots \pi(f-1)=f,\pi(f)=1$, then we have
\begin{equation}
  B_{\sigma} = (-1)^{f-1}e^{\i \phi}z_{\sigma(1)}^N B_{\sigma\cdot \pi}.
  \label{eqn:cyclicshift}
\end{equation}

\paragraph{Bethe equations.} The cyclic shift property allows to derive the Bethe equations for the model. For $f$ particles the amplitudes $B_\sigma$ are required to solve the system of equations 
\begin{equation}
  B_{\cdots \sigma(i),\sigma(i+1),\cdots} r(u_{\sigma(i)},u_{\sigma(i+1)})+B_{\cdots \sigma(i+1),\sigma(i),\cdots} r(u_{\sigma(i+1)},u_{\sigma(i)})=0, \quad \sigma \in S_f.
  \label{eqn:betheequations}
\end{equation}
where $r(u,v)$ is the function defined in \eqref{eqn:rofuandv}. The system can be solved by
\begin{equation*}
  B_\sigma = C^{-1}\,\text{sgn}\,\sigma \prod_{1\leq m<n\leq f}b_{\sigma(m)\sigma(n)}.
\end{equation*}
Here $C$ is an arbitrary normalisation factor, and the numbers $b_{mn}$ are solutions to the equations
\begin{equation*}
  b_{mn} r(u_m,u_n) = b_{nm} r(u_n,u_m).
\end{equation*}
If all the $r(u_m,u_n)$ are finite and non-zero for all $m,n$ then this system of equations has the simple solution $b_{mn} = r(u_n,u_m)$. Using the cyclic shift property, we obtain in this case the Bethe equations of the model:
\begin{equation}
  z(u_k)^{N-f} = e^{-\i \phi}\prod_{j=1}^f \frac{\vartheta_1(u_j-\theta)}{\vartheta_1(u_j+\theta)}\frac{\vartheta_1(u_k-u_j-\theta)}{\vartheta_1(u_k-u_j+\theta)}
  \label{eqn:BAE}
\end{equation}
A solution of this equation leads to an eigenstate of the Hamiltonian whose energy is given by
\begin{equation}
  E= N+\sum_{j=1}^f \epsilon(u_j)
  \label{eqn:totalenergy}
\end{equation}
where $\epsilon(u)$ is the elliptic form of the excitation energy found in \eqref{eqn:excitationenergy}. In particular, as the Bethe equations do not depend on the parameter $t$ which parametrises the constants $\mu_x,\lambda_x$, nor does the excitation energy itself, we conclude that the energy is $t$-independent. It follows that the spectrum does not change as one moves along the constant $q$ lines on the special submanifold in parameter space plotted in figure \ref{fig:tparam}. The eigenstate itself can be reconstructed from the amplitudes $B_\sigma$. With an appropriate choice of normalisation, we find that they are given by
\begin{equation*}
  B_\sigma = \text{sgn}\,\sigma \prod_{n=1}^f z(u_{\sigma(n)})^{-n} \prod_{1\leq m < n \leq f}\vartheta_1(u_{\sigma(n)}-u_{\sigma(m)}+\theta).
\end{equation*}

There are however cases, where $r(u_m,u_n)$ vanishes or becomes infinite for certain pairs $m,n$. This happens for so-called exact strings or bound states, which need to be treated separately. We will show in the next section, that these somewhat exceptional cases are actually quite relevant in order to understand the supersymmetry from the point of view of the Bethe ansatz.

\section{Supersymmetry and the Bethe ansatz}
\label{sec:bethesusy}

In this section we analyse the relation between the different symmetries of the model and the Bethe ansatz equations. We show that the action of the operators $Q_+$ and $Q_0$ can be derived rather straightforwardly from the Bethe equations. The dynamic supersymmetry generated by $Q_-$ is however more subtle. As we shall see it is related to the existence of so-called exact strings of Bethe roots, which are present in the model essentially because the parameter $\theta$ is a rational multiple of $\pi$.

\subsection{Non-dynamic supersymmetry $Q_+$}
Let us start with the supersymmetry that was originally used to define the model. We claim that the action of $\bar Q_+$ is equivalent to adding to a set of Bethe roots $u_1,\dots, u_f$ that solve \eqref{eqn:BAE} an additional root $u_{f+1}=0$, i.e. rapidity $z_{f+1}=1$, without changing the number of sites. This is readily verified by comparing the Bethe equations at $f$ and $f+1$ particles, which confirms our statement provided that the twist angle is $\phi=0$. The relation between the wave functions with $f$ and $f+1$ particles can be evaluated explicitly for $z_{f+1}=1$:
\begin{equation*}
  \psi(x_1,\dots,x_{f+1}) = \text{const.} \times \sum_{k=1}^{f+1}(-1)^{k-1}\lambda_{x_k}\psi(x_1,\dots,x_{k-1},x_{k+1},\dots,x_{f+1})
\end{equation*}
The string $(-1)^{k-1}$ is a clear sign of a fermionic operator. Promoting this relation between wave functions to a relation between the corresponding states by multiplying each side with $||x_1,\dots,x_{f+1}\rrangle$, followed by a summation over all allowed particle arrangements, leads straightforwardly to the definition of $\bar Q_+$ (up to a constant).

\subsection{Dynamic supersymmetry $Q_0$} The neutral dynamic supersymmetry can also be understood through a simple addition of a Bethe root to a given solution $u_1, \dots, u_f$ of \eqref{eqn:BAE}. The new member has $u_{f+1}= \pi/2$, and corresponds therefore to a particle with rapidity $z_{f+1}=-1$. In addition to this particle insertion, one needs to increase the length of the chain by two.

From the Bethe ansatz form of the wave function we obtain a relation between the wave functions for $f+1$ particles with $z_{f+1}=-1$, and $f$ particles with arbitrary rapidities:
\begin{equation*}
  \psi(x_1,\dots,x_{f+1}) = \text{const.} \times \sum_{k=1}^{f+1}(-1)^{x_k+k-1}\lambda_{x_k+1}\psi(x_1,\dots,x_{k-1},x_{k+1}-2,\dots,x_{f+1}-2)
\end{equation*}
It is not difficult to translate this equation into a relation between the corresponding eigenvectors of the Hamiltonian at $N$ and $N+2$ sites. The corresponding operator is fermionic and carries ``momentum'' $\pi$ as can be seen from the string $(-1)^{x_k+k-1}$. Working out its amplitudes leads precisely to the dynamic supercharge $Q_0$ discussed in section \ref{sec:dynneutralsusy}.

\subsection{Dynamic supersymmetry $Q_-$} The Bethe equations for the staggered $M_2$ model resemble those of the eight-vertex model at so-called root-of-unity points, i.e. points where $\theta$ is a rational multiple of $\pi$. It is known that at such points so-called exact strings appear in finite-size systems, i.e. configurations of Bethe roots which are arranged in the pattern
 \begin{equation}
 \label{eqn:exactstringbetheroots}
 u_j = u + (j-1) \theta, \qquad \theta=\pi/4,
 \end{equation}
with $j=1, 2, 3, 4$. The aim of this section is to discuss a relation between these exact strings and the dynamic supercharge $Q_-$. To this end, we derive the wave function for a single exact string and then relate it to the action of $Q_-$ and $\bar{Q}_+$ in the limit where the so-called string centre $u$ tends to zero. 

Let us first discuss a few properties of an exact string of Bethe roots. From the elliptic parameterisation of the Bethe roots given in \eqref{eqn:rapidityparam}, we infer that its total rapidity is given by
\begin{equation*}
  \prod_{j=1}^4 z(u_j) = 1,
\end{equation*}
and hence does not carry any net momentum.
Moreover, the sum of the single particle excitation energies for its members yields
\begin{equation*}
  \sum_{j=1}^4 \epsilon(u_j)=-4,
\end{equation*}
irrespectively of the value $u$ for the string centre.
Comparing this with the expression of the total energy in \eqref{eqn:totalenergy}, we conclude that adding an exact string to a configuration of Bethe roots decreases the energy by four. Therefore, if we add simultaneously four sites to the system, the total energy remains unchanged. This observation hints at a dynamic symmetry relating the Hamiltonians for chains of length $N$ and $N+4$.

Here, we investigate the simplest case of a single exact string in order to establish a relation with dynamic supersymmetry. To this end, we compute the wave function by following Baxter's calculation for the six-vertex model \cite{baxter:02}. Concretely, for four particles we have to solve the equations, \eqref{eqn:betheequations},
\begin{equation*}
B_{\cdots \sigma(i),\sigma(i+1),\cdots} r(u_{\sigma(i)},u_{\sigma(i+1)})+B_{\cdots \sigma(i+1),\sigma(i),\cdots} r(u_{\sigma(i+1)},u_{\sigma(i)})=0.
\end{equation*}
These give the relations between the different amplitudes of the wave function, $B_\sigma$, where $\sigma \in S_4$ in the present case. For the configuration of Bethe roots that form the exact string \eqref{eqn:exactstringbetheroots} it can easily be seen that the function $r(u_i, u_j)$ defined in \eqref{eqn:rofuandv} vanishes whenever $j=i+1$ for $i=1,2,3$, or $i=4, j=1$. Conversely, $r(u_j, u_i)$ is {\emph nonzero} for these cases. If we normalise the amplitudes such that $B_{1234}\neq 0$ then our equations imply that all other amplitudes are finite, and that $B_{2134}=0$ because $r(u_2, u_1)=0$, $B_{1324}=0$ because $r(u_3, u_2)=0$ etc. We find that all $B_{\sigma}$ are zero except those for which the permutation $\sigma$ is an integer power of the cyclic shift $\pi =(1234)$. The remaining amplitudes are related because of the translation symmetry \eqref{eqn:cyclicshift}. For zero twist angle, we find the simple relation
\begin{equation*}
  B_{j+1, \ldots, 4,1,\ldots, j}=(-1)^j z_{j+1}^N \ldots z_{4}^N B_{1234}.
\end{equation*}

We obtain the wave function for a single exact string by plugging into the Bethe wave function, \eqref{eqn:betheansatz}, the expression for the non-vanishing $B_\sigma$. Writing out the rapidities in their elliptic parametrisation, we find
\begin{equation*}
\psi(x_1,x_2,x_3,x_4) = C \sum_{j=0}^{3} (-1)^{x_1 + x_2} \prod_{k=1}^{4} A_{x_k} \vartheta_1 (u + (k+j-2) \theta)^{x_{k+2}-x_k -3},
\end{equation*}
where
\begin{equation*}
  A_{x_k}= \lambda_{x_k}  \frac{\vartheta_4 (t-u_1 + (2x_k - (k+j))\theta)}{\vartheta_4(t+(2x_k -1) \theta)}
\end{equation*}
and $C$ is some normalisation constant.

We now take the limit where the centre of the exact string $u$ tends to zero and establish the relation with the supercharge $Q_-$. To this end, observe that for $u=0$ the wave function is only nonzero for a special set of configurations. In this case, the products in the exact-string wave function contain a factor $\vartheta_1((k+j-2)\theta)$ which is zero when $k=2-j,\,6-j$, so for every $j$ there is a case for which this becomes zero. The whole wave function vanishes therefore unless also $x_{k+2}-x_k -3=0$ simultaneously. This means that the fermion configuration contains a pair of particles located at $x_k$ and $x_{k+2}$, such that $x_{k+2}-x_k = 3$. Now, recall that because the particles are ordered there needs to be one particle on the two sites in between them. This fixes the relative positions of three particles. The remaining one is itinerant: it can be anywhere as long as the exclusion constraints of the model hold. For simplicity, we consider only the configurations for which the first site is occupied. The non-vanishing values of the Bethe wave functions for a chain of $N$ sites are given by:
\begin{align*}
& \underset{1}{\bullet}\circ \underset{3}{\bullet}\underset{4}{\bullet}\circ \cdots \underset{x}{\bullet} \cdots \underset{N}{\circ} \\
&\qquad \psi(1,3,4,x)=C (\vartheta_1(\theta) \vartheta_1(2 \theta))^{N-6} \lambda_1^2 \lambda_0 \lambda_x \frac{\vartheta_4(t)^2}{\vartheta_4(t+\theta)^2}\\
&\underset{1}{\bullet} \underset{2}{\bullet} \circ \underset{4}{\bullet} \circ \cdots \underset{x}{\bullet} \cdots \circ \underset{N}{\circ}\\
& \qquad \psi(1,2,4,x)=-C(\vartheta_1(\theta) \vartheta_1(2 \theta))^{N-6}\lambda_0^2 \lambda_1 \lambda_x \frac{\vartheta_4(t)^2}{\vartheta_4(t+3\theta)^2} \\
&\underset{1}{\bullet} \circ \underset{3}{\bullet} \underset{4}{\bullet} \circ \underset{6}{\bullet} \circ \cdots \underset{N}{\circ} \\ 
& \qquad \psi(1,3,4,6)=C(\vartheta_1(\theta) \vartheta_1(2 \theta))^{N-6} \vartheta_4(t)^2 \lambda_0^2 \lambda_1^2  \left(\frac{1}{\vartheta_4(t+3\theta)^2}+\frac{1}{\vartheta_4(t+\theta)^2}\right)\\
&\underset{1}{\bullet} \underset{2}{\bullet} \circ \underset{4}{\bullet} \underset{5}{\bullet}\circ \cdots \underset{N}{\circ}\\
 & \qquad \psi(1,2,4,5)=-C(\vartheta_1(\theta) \vartheta_1(2 \theta))^{N-6} \lambda_1^2 \lambda_2^2 \left(\frac{\vartheta_4(t)^2}{\vartheta_4(t+3 \theta)^2} + \frac{\vartheta_4(t+2 \theta)^2}{\vartheta_4(t+3\theta)^2}\right).
\end{align*}
All other non-vanishing amplitudes can be recovered from these either from invariance under translation by two sites, or by shifting $t\to t + 2\theta$, which amounts to a translation by one site. We now normalise the amplitudes by setting
\begin{equation*}
  C=-\frac{1}{(\vartheta_1(\theta) \vartheta_1(2 \theta))^{N-6} } \frac{\vartheta_4(t+\theta)^2}{\vartheta_4(t)^2}\frac{1}{\lambda_1^2} \frac{\mu_0}{\mu_1}.
\end{equation*}  
The Bethe wave function gives the amplitudes of the non-orthonormal basis we defined in \eqref{eqn:nonorthonormalbasis}. To find the amplitudes of the configurations in the basis in which we defined $Q_-$ \eqref{eqn:localaction} we  have to include an extra normalisation factor $\mu_x$ for any pair starting a site $x$. When comparing with the definitions of the non-dynamic and dynamic supercharges in section \ref{sec:defmodel} and \ref{sec:dynsusy} we observe that we obtain after some algebra precisely the amplitudes of $Q_-\bar{Q}_+(=-\bar{Q}_+Q_-)$ acting on the empty chain:
\begin{align*}
& \mu_3\psi(1,3,4,x) =-\mu_0\lambda_0 \lambda_x ,\quad  
\mu_1\psi(1,2,4,x)=\mu_0\lambda_1  \lambda_x,\quad\text{for}\quad 5<x<N,\\
& \mu_3\psi(1,3,4,6) =-2\mu_0, \quad
\mu_1\mu_4\psi(1,2,4,5)=\lambda_1^2.
\end{align*}

Remember that we showed that $\bar{Q}_+$ acts on Bethe states by adding a Bethe root $u=0$ to a configuration. We conclude that $Q_-$ adds a set of three Bethe roots which is centred around the Bethe root $u=0$, in a pattern of an exact string with exactly the central root missing. To be more precise, the above argument only shows this for $Q_-$ acting on an empty chain, but we expect the action to be the same when starting from a general configuration of Bethe roots. Another possible way to prove this statement might be to study the insertion of the three Bethe roots $u=\pi/4,\pi/2,3\pi/4$ to a given solution of the Bethe equations, and increase the number of sites by four. However, it appears that this has to be done by employing a suitable limiting procedure which involves breaking translation symmetry, and is therefore technically very challenging.


\section{Ground states}
\label{sec:cohomology}

In this section we analyse the zero-energy states of the model. We determine their number for a given system size from cohomological arguments in section \ref{sec:cc}. In section \ref{sec:gsba} we find their Bethe roots in finite size from a set of functional equations.

\subsection{Cohomology computation of the number of ground states}
\label{sec:cc}

In supersymmetric theories the zero-energy ground states enjoy special properties since they are singlet representations of the supersymmetry algebra. A lower bound on the number of zero energy states is given by the absolute value of the Witten index \cite{witten:82}. The Witten index is defined as $W=\textrm{Tr} (-1)^F e^{-\beta H}$, where the trace is over the entire Hilbert space. Since all positive energy states come in pairs that differ in their fermion number by one, these states do not contribute to the Witten index. We can thus take the limit $\beta \to \infty$ and write $W=\textrm{Tr} (-1)^F$ where the trace is now restricted to the space spanned by the zero-energy states. Clearly, this index is independent of the parameters of the Hamiltonian (provided the Hilbert space remains unchanged). It follows that the Witten index of the inhomogeneous $M_2$ model is given by the Witten index of the homogeneous $M_2$ model, which was computed in \cite{fendley:03}.

In this section we present the natural extension to this result, which is the computation of the exact number of zero energy states and the number of fermions in the zero energy states. To this end, we treat in addition to periodic boundary conditions the case of so-called open boundary conditions. These correspond simply to considering a linear chain without identifying the first and last site as neighbours. The definition of the supercharge $Q_+$ on these chains is straightforward. For the number of ground states of the model with open and periodic boundary conditions, we obtain the following result:

\begin{theorem}\label{thm:GS} Consider the  $M_2$ model with non-vanishing weights $\lambda_x$ and $\mu_x$. For open boundary conditions the model has exactly one zero-energy state in $\mathcal H_f$ with $f=2n$ if the length of the chain is $N=4n$ or $N=4n-1$, and none otherwise. For periodic boundary conditions there are $3$ zero-energy states in $\mathcal H_f$ with $f=2n$ if $N=4n$, and a single zero-energy state otherwise with $f=\lfloor N/2 \rfloor=2n,2n+1,2n+2$ for $N=4n+1,4n+2,4n+3$, respectively.
\end{theorem}

It is important to impose the condition that the staggering parameters are non-vanishing. When the weights are allowed to be zero the number of zero-energy state may increase dramatically, as discussed in \cite{fendley:03,beccaria:12,huijse:11}. 

To prove the theorem, we use the fact that zero-energy states are in one-to-one correspondence with the cohomology elements of the supercharges. For a more leisurely explanation of this relation we refer the reader to \cite{huijse:10}. Here we merely state that, together with the supercharge, the Hilbert space forms a chain complex, where the fermion number provides the grading, $\bar{Q}_+ : \mathcal{H}_{N,f} \to \mathcal{H}_{N,f+1}$. The cohomology of $\bar{Q}_+$ at grade $f$ is defined as $\mathfrak H^{(f)}= \textrm{Ker}(\bar{Q}_+)/\textrm{Im}(\bar{Q}_+)$ within $\mathcal{H}_{N,f}$. Roughly speaking, a state is in the cohomology of $\bar{Q}_+$ when it is annihilated by $\bar{Q}_+$, but cannot be written as $\bar{Q}_+$ of something else. Note that since zero-energy states are singlets of the superalgebra, they precisely obey this condition. Finally, to compute the cohomology of the supercharge, we use the `tic-tac-toe' lemma \cite{BottTu82}. This lemma says that if we define $\bar{Q}_+$ as the sum of two operators acting on two disjoint sublattices, $S_1$ and $S_2$, the cohomology of $\bar{Q}_+$ isomorphic to $\mathfrak H_{12}=\mathfrak H_1 (\mathfrak H_2)$, where $\mathfrak H_i$ is the cohomology of the supercharge acting on sublattice $S_i$, provided that $\mathfrak H_{12}$ contains non-trivial elements only at one grade, i.e. one value of $f$. We will see below, that for a clever choice of the sublattices, this is always the case here.

We now proceed to the proof of theorem \ref{thm:GS}. The basic idea is to compute the cohomology of the supercharges in two steps using the spectral sequence technique. To this end, the lattice is divided into two disjoint parts $S_1$ and $S_2$. The first step consists of evaluating the cohomology for the supercharge restricted to $S_1$. Within the resulting space we then evaluate the cohomology of the supercharge restricted to $S_2$.
  
  \bigskip
  
  We start with \textbf{open boundary conditions}. We choose for $S_2$ the sites $1,5,\dots,4n+1$ where $n$ is defined through $N=4n+p,\,p=1,\dots,4$. $S_1$ is defined as the collection of the remaining sites. Consider the supercharge $\bar{Q}_{+,2} =\bar{Q}_{+}|_{S_2}$ which is the restriction of $\bar{Q}_+$ to $S_2$. Any cohomology element, or cycle, of $\bar{Q}_{+,2}$ is necessarily represented by a state of the form
  \begin{equation*}
    |{\underset{\uparrow}{\circ}\bullet\bullet\circ\underset{\uparrow}{\circ}\bullet\bullet\circ\cdots \underset{\uparrow}{\circ}\bullet\bullet\circ\hspace{.06cm}\underset{\uparrow}{\circ}}\rangle \otimes |\psi\rangle
  \end{equation*}
where the arrows indicate the sites of $S_2$. Here, $|\psi\rangle$ is a state with $p-1$ sites. To see this, first note that all the sites of $S_2$ have to be empty otherwise the state is in the image of $\bar{Q}_{+,2}$. Furthermore, for the state to be in the kernel of $\bar{Q}_{+,2}$ each site of $S_2$ needs to be adjacent to a pair of particles. If we look at the site $1\in S_2$, this implies that a pair has to be present on sites $2$ and $3$. Now, the exclusion rule implies that site $4$ has to be empty. Next, if we look at the site $5 \in S_2$, the problem is identical to the previous one, hence there has to be a pair on sites $6$ and $7$. By recursion we thus arrive at site $4n+1 \in S_2$. Again, it has to be adjacent to a pair. For $p<3$ the problem has no solution. If $p=3,4$ the solution is unique:
  \begin{equation*}
    |\psi\rangle =
    \begin{cases}
       |\bullet\bullet\rangle ,& p = 3,\\
       |\bullet\bullet\hspace{.06cm}\circ\rangle,& p = 4.
    \end{cases}
  \end{equation*}
  
We conclude that the cohomology $\mathfrak H_{2}$ of $\bar{Q}_{+,2}$ has dimension one if $N=4n$ or $N=4n-1$, and zero otherwise. Note that the total number of particles is $f=2n$. Now we act with $\bar{Q}_{+,1}$ on $\mathfrak H_{2}$. When $N=4n+1$ or $N=4n+2$ the dimension of  $\mathfrak H_{2}$ is zero and therefore the dimension of $\mathfrak H$ is zero. When  $N=4n$ or $N=4n-1$ then $\mathfrak H_{2}$ contains one element, it follows directly that this element cannot be in the image of $\bar{Q}_{+,1}$ and has to be in the kernel of $\bar{Q}_{+,1}$. We thus find that for $N=4n$ or $N=4n-1$, the dimension of  $\mathfrak H^{(2n)}$ is one and it is zero at all other grades.
  
  \bigskip
  
  Next, we consider \textbf{periodic boundary conditions} for chains with length $N=4n$. As before, we choose for $S_2$ the sites $1,5,\dots,4n-3$. For a state to be in the kernel but not in the image of $\bar{Q}_{+,2}$ it should have all sites of $S_2$ empty, furthermore each such site must be adjacent to two particles. Pick an arbitrary site of $S_2$, then this condition corresponds to
  \begin{equation*}
    \cdots \circ \underset{\uparrow}{\circ} \bullet\bullet\circ\cdots,\quad \textrm{or} \quad 
 \cdots \circ\bullet \underset{\uparrow}{\circ} \bullet\circ\cdots,\quad \textrm{or}\quad
 \cdots \circ\bullet\bullet \underset{\uparrow}{\circ} \circ \cdots.
  \end{equation*}
  By recursion, one shows easily that this can hold for all sites of $S_2$ if and only if the particle distribution in their immediate neighbourhood is the same for all of them. This leads to a total of $3$ cycles of $\bar{Q}_{+,2}$. Notice that all of them contain exactly $f=2n$ particles, this implies immediately that within $\mathfrak H_2$ all states are in the kernel of $\bar{Q}_{+,1}$ and none are in the image of $\bar{Q}_{+,1}$. We conclude that for $N=4n$ the dimension of  $\mathfrak H^{(2n)}$ is $3$ and it is zero at all other grades.
  
  \bigskip
  
It remains to prove the theorem for chains with length $N\neq 4n$ and \textbf{periodic boundary conditions}. We first consider $N=4n+1$ and take $S_1$ to be two consecutive sites and $S_2$ the rest of the chain. When $S_1$ is empty, $\bar{Q}_{+,2}$ acts on an open chain of length $4n-1$. Using our previous results, we find that the cohomology of $\bar{Q}_{+,2}$ is one-dimensional. Similarly, when both $S_1$ sites are occupied, $\bar{Q}_{+,2}$ acts on an open chain of length $L=4n-3$ because both sites on $S_2$ that are adjacent to the $S_1$ sites have to be empty. It follows that the cohomology of $\bar{Q}_{+,2}$ vanishes when both $S_1$ sites are occupied. Finally, we must consider the case where one of the $S_1$ sites is occupied and the other is empty. The cohomology problem of $\bar{Q}_{+,2}$ in this case remains to be solved. We now have that $S_2$ is an open chain of length $4n-1$, but with an unusual boundary condition at one end, namely the last site can be occupied, but only if the penultimate site is empty, that is the last site cannot be part of a pair. We use another spectral sequence to address this problem. Let us label the sites of $S_2$ by $i=1, 2,\dots, 4n-1$ and the $S_1$ sites are the sites $0$ and $4n$. Furthermore, consider the case that site $0$ is empty and site $4n$ is occupied. Now take $S_A$ to be the sites $1,5,\dots,4n-3$ and $S_B$ the rest of the $S_2$ sites: $S_B=S_2 \backslash S_A$. It is now easily seen that the cohomology of $\bar{Q}_{+,A}$ vanishes: the site $0$ is empty and therefor we need sites 2 and 3 to be occupied for $\bar{Q}_{+,A}$ to vanish on the first site. Continuing this argument we find that we also need to occupy sites $4n-2$ and $4n-1$, but that is not allowed since site $4n$ is already occupied. Since $\mathfrak H_{A}$ is zero dimensional it follows that  $\mathfrak H_{B}(\mathfrak H_{A})$ vanishes. We thus conclude that $\mathfrak H_{2}$ is also empty when one of the $S_1$ sites is occupied and the other is empty. We have thus found that $\mathfrak H_{2}$ is one-dimensional: 
  \begin{equation*}
    |\circ\circ\rangle_{S_1} \otimes |\psi  \rangle_{S_2},
  \end{equation*}
with $|\psi \rangle$ the unique cohomology element of the supercharge for the open chain of length $4n-1$. Computing the cohomology of $\bar{Q}_{+,1}$ within $\mathfrak H_{2}$ is then trivial and thus we can conclude that for $N=4n+1$ the dimension of $\mathfrak H^{(2n)}$ is one and the cohomology is trivial at all other grades. 

In a very similar manner we can prove that $\mathfrak H^{(2n+2)}$ is one-dimensional for $N=4n+3$ and trivial at all other grades. It is clear that both $S_1$ sites empty is not an element of the cohomology of $\bar{Q}_{+,2}$ and both $S_1$ occupied is. When one of the $S_1$ sites is empty and the other is occupied, we can proceed as above. One can easily verify that the cohomology of $\bar{Q}_{+,A}$ vanishes by taking $S_A=\{i|i=1,5,\dots,4n+1\}$. 

Finally, we also find that $\mathfrak H^{(2n+1)}$ is one-dimensional for $N=4n+2$ and  trivial at all other grades by taking $S_1$ to be a single site. When $S_1$ is empty the cohomology of $\bar{Q}_{+,2}$ vanishes, however, when $S_1$ is occupied we do find a non-trivial element. This can be seen by taking $S_1$ to be the site zero and taking $S_A=\{i|i=1,5,\dots,4n+1\}$. The only solution is
  \begin{equation*}
    |\bullet\rangle_{S_1} \otimes |\underset{\uparrow}{\circ} \bullet\circ\bullet \underset{\uparrow}{\circ} \bullet\circ\bullet \cdots \underset{\uparrow}{\circ} \bullet\circ\bullet \hspace{.06cm}\underset{\uparrow}{\circ}  \rangle_{S_2},
  \end{equation*}
where the arrows indicate the sites of $S_A$. This completes the proof of theorem \ref{thm:GS}.

\subsection{Ground states and the Bethe ansatz}
\label{sec:gsba}

The existence of exact finite-size zero-energy states for periodic boundary conditions suggests that it might be possible to determine their Bethe roots exactly. The purpose of this section is to show that this is indeed the case. We introduce an analogue of the $\mathcal T$-$\mathcal Q$ equation for our model. The analogy is of course formal because unlike for integrable vertex models, we do not dispose of a transfer matrix in the present setting. Nonetheless, using an analyticity argument we show that the formal analogy is sufficient to find the ground-state Bethe root distribution from certain functional equations. Throughout this section we consider periodic boundary conditions: the twist angle is zero, $\phi=0$.

\subsubsection{The functional equations}
Let us introduce a function whose roots coincide with the Bethe roots for a chain with $N$ sites and $f$ particles:
\begin{equation}
  \mathcal Q(u) = \prod_{j=1}^f \vartheta_1(u-u_j).
  \label{eqn:qfunction}
\end{equation}
Furthermore, inspired by the parallels between our model and integrable quantum spin$-1$ chains as explained in section \ref{sec:susy}, we introduce a system of functional equations akin to the fusion equations for the nineteen-vertex model (see for example \cite{kirillov:87}):
\begin{align*}
  \mathcal T_2(u) &= \mathcal T_1(u)\mathcal T_1(u+\theta) + (-1)^N \mathfrak t\Phi(u)\Phi(u+\theta),\\
  \mathcal T_1(u)\mathcal Q(u) &= \Phi(u+\theta)\mathcal Q(u+\theta) + (-1)^{N+1}\mathfrak t \Phi(u-\theta)\mathcal Q(u-\theta).
\end{align*}
Here, $\mathcal T_1(u),\mathcal T_2(u)$ are unknown functions, $\Phi(u) = \vartheta_1(u)^{N-f}$, and $\mathfrak t = (-1)^f \mathcal Q(\theta)/\mathcal Q(-\theta)$ is a number whose square coincides with the eigenvalue of the translation operator by two sites \eqref{eqn:eigenvaluetransop}. The second equation corresponds to Baxter's $\mathcal T$-$\mathcal Q$ relation. If we impose that $\mathcal T_1(u)$ be analytic in $u$ then it implies the Bethe equations: indeed, the analyticity requirement means that the left-hand side has no poles. Setting thus $u=u_k$ for $k=1,\dots,f$ makes it vanish, and imposing the same on the right-hand side leads then to the Bethe equations \eqref{eqn:BAE} for zero twist angle.
Furthermore, combining both equations it is not very difficult to show that the function $\mathcal T_2(u)$ has the following properties
\begin{equation}
  \mathcal T_2(\theta) = \mathfrak t^{-1} \Phi(\theta)\Phi(2\theta), \quad \frac{\mathcal T_2'(\theta)}{\mathcal T_2(\theta)} = \frac{\vartheta_1'(\theta)}{\vartheta_1(\theta)}(N-f-2E).
  \label{eqn:propertiesT2}
\end{equation}

Our aim is therefore to determine the solution to the functional equations for the supersymmetry singlets. We will argue that it is given by
\begin{equation*}
  \mathcal T_1(u) = \eta \Phi(u).
\end{equation*}
where $\eta$ is a constant to be determined.
Indeed, it is easily shown that this choice implies \eqref{eqn:propertiesT2} with $E=0$, provided that
\begin{equation}
  \eta^2 = \mathfrak t^{-1}+(-1)^{N+1}\mathfrak t.
  \label{eqn:constrainteta}
\end{equation}
Moreover, this ansatz is analytic in the variable $u$, and therefore implies the Bethe equations. Furthermore, the insertion of the ansatz into the $\mathcal T$-$\mathcal Q$ equation fixes the fermion number $f$ for given $N$ to the values which we found from cohomology, as we shall see in the following two subsections. Before proceeding, let us however point out a caveat: finding a solution to the Bethe equations with energy $E=0$ leads only to a supersymmetry singlet if the corresponding Bethe wave function is non-vanishing. Even though this is readily checked numerically for the solutions we present here for small system size, $N$, a general proof appears to be difficult, and therefore our results remain conjectural.

\subsubsection{Solution of the functional equations in the trigonometric limit} 
As the results in the general elliptic case is rather complicated, we start with a discussion of the trigonometric limit where the elliptic nome tends to zero $q\to 0$, and the model becomes translation invariant. In this limit $\mathfrak t$ is simply the eigenvalue of the translation operator. Furthermore, the $\mathcal Q$-function \eqref{eqn:qfunction} vanishes in this limit according to
\begin{equation*}
  \mathcal Q(u) = (2q^{1/4})^f\mathcal Q_h(u)+\dots,\qquad \mathcal Q_h(u) = \prod_{j=1}^f \sin(u-u_j).
\end{equation*}
In order to simplify matters, we introduce the function $\mathcal R(u) = (\sin u)^{N-f} \mathcal Q_h(u)$. The trigonometric limit of the $\mathcal T$-$\mathcal Q$ equation can then conveniently be rewritten as
\begin{equation}
  \eta\;\mathcal R(u) = \mathcal R(u+\theta) + (-1)^{N+1}\mathfrak t\; \mathcal R(u-\theta).
  \label{eqn:recursionR}
\end{equation}
Clearly $\mathcal R(u)$ is a trigonometric polynomial of degree $N$, and we may write $\mathcal R(u) = e^{-\i N u} \sum_{j=0}^N c_j e^{2\i j u}$. The equation for $\mathcal R(u)$ leads then to the following equation
\begin{equation*}
  (\eta-e^{\i(2j- N) \theta}-(-1)^{N+1} e^{\i(N-2j)\theta}\mathfrak t )c_j=0,
\end{equation*}
which implies that the coefficient $c_k$ vanishes unless the prefactor within the brackets is zero. We impose $c_0,c_N \neq 0$, which leads to the following two requirements:
\begin{align*}
  \eta = e^{-\i N \theta}+ (-1)^{N+1} e^{\i N \theta}\mathfrak t =e^{\i N \theta}+ (-1)^{N+1} e^{-\i N\theta}\mathfrak t.
\end{align*}
The two expressions for $\eta$ need to be compatible. Furthermore, if this condition is met we need to make sure that \eqref{eqn:constrainteta} holds. These two constraints restrict $\mathfrak t$ to certain admissible values.

\paragraph{Chains of length $\mathbf{N=4n}$.} Let us start the analysis with the case where the number of sites is a multiple of four: $N=4n$. We find 
\begin{align*}
  \mathfrak t = - e^{2\i m \theta},\, \eta = (-1)^n(1+e^{2\i m \theta}), \quad m=1,2,3.
\end{align*}
We insert these values the in functional equation for $\mathcal R(u)$ and use its explicit form. A short calculation then shows that we need to set $c_j=0$ unless $j=4k,4k+m$ where $k$ is integer. The remaining coefficients are determined from the requirement that by definition $\mathcal R(u)$ has a zero of order $N-f$ at $u=0$, which implies $\mathcal R(0) = \mathcal R'(0)=\cdots=\mathcal R^{(N-f-1)}(0)=0$, and therefore $4n-f$ homogeneous linear equations for $2n+1$ unknowns. This gives a non-trivial solution if the number of constraints exceeds the number of unknowns by one, i.e. for half-filling $f=2n = N/2$. In this case, the function $\mathcal R(u)$ can be computed in terms of hypergeometric functions by following the strategies of \cite{fridkin:00,fridkin:00_2}. Up to an unimportant factor, the exact result is 
\begin{align}
  \mathcal R(u) =& e^{-\i N u}\Biggl({_2}F_1\Bigl(-n,-n+1-\frac{m}{4},1-\frac{m}{4};e^{8\i u}\Bigr) \label{eqn:RN4n}\\
    &-\frac{\Gamma\left(1-\frac{m}{4}\right)}{\Gamma\left(1+\frac{m}{4}\right)} \frac{n \Gamma\left(\frac{m}{4}-n\right)}{\Gamma\left(1-\frac{m}{4}-n\right)}e^{2\i m u}{_2}F_1\left(-n+1,-n+\frac{m}{4},1+\frac{m}{4};e^{8\i u}\right) \Biggr)  \nonumber
\end{align}
for $m=1,2,3$ and $N=4n$.

The explicit solution may be used in order to characterise the distribution of the Bethe roots in the complex plane. Let us first note that a drastic simplification takes place in the case $m=2$, i.e. $\mathfrak t=1$. In fact, the special functions simplify to $\mathcal R(u) = (-2\i \sin 2u)^{2n}$. Hence, the corresponding $\mathcal Q$-function has a zero of order $f=2n$ at $u= 2\theta = \pi/2$, and thus coinciding Bethe roots. While the Bethe ansatz wave function vanishes naively when setting all Bethe roots to the same value one may consider the limit where they tend to this value, and renormalise it properly so that the result is finite. We checked that this procedure does indeed give a zero-energy eigenstate of the Hamiltonian, and therefore conjecture the following:
\begin{conjecture}
The ground state in the sector $f=N/2$ and $\mathfrak t=1$ for $N=4,8,\dots$ for periodic boundary conditions is the limit of an appropriately renormalised Bethe state where all Bethe roots $u_1, u_2,\dots, u_f$ tend to $u=2\theta$. 
\label{conj:condensation1}
\end{conjecture}

\begin{figure}[h]
  \centering
  \includegraphics[scale=0.45]{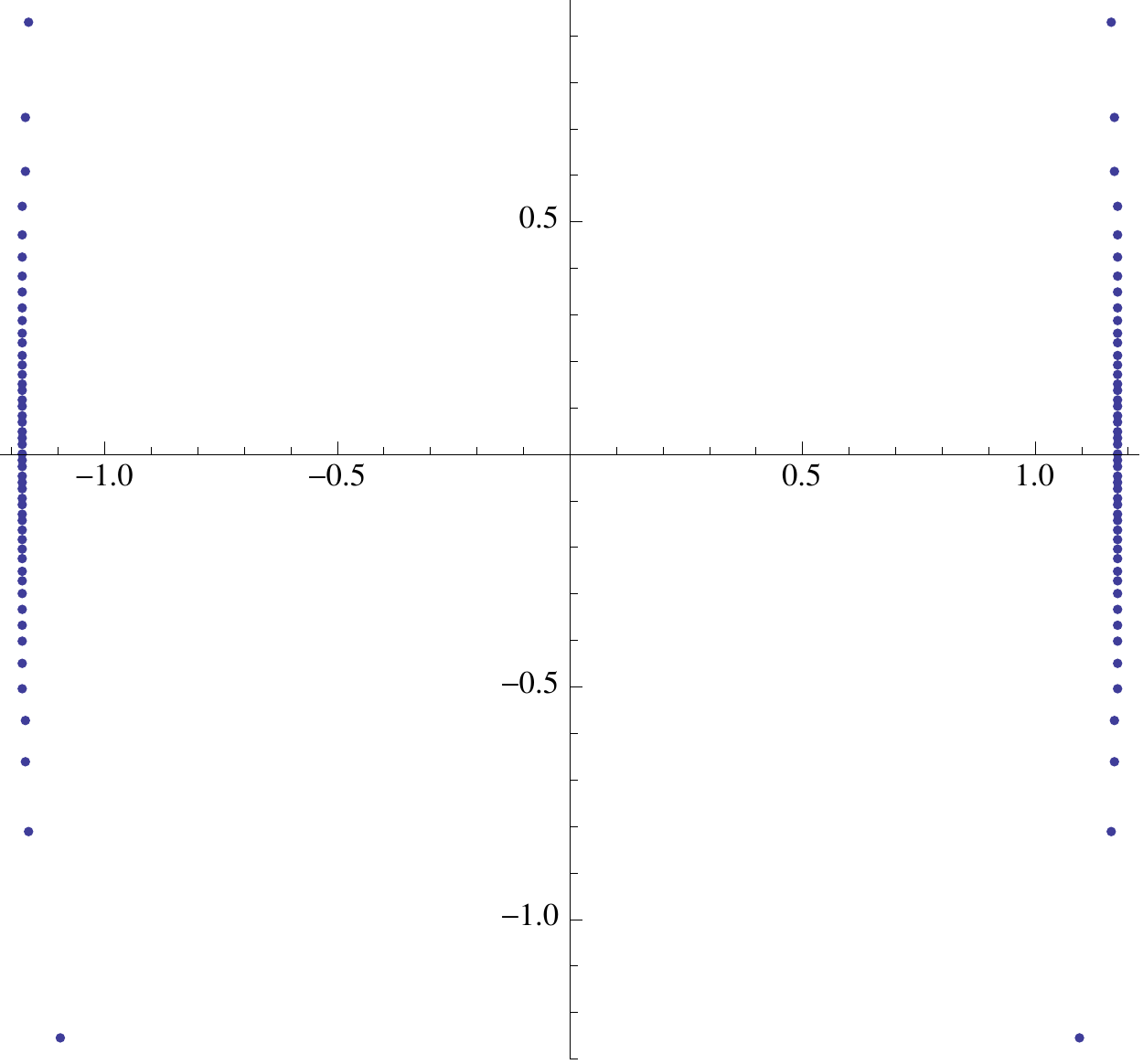}
  \includegraphics[scale=.55]{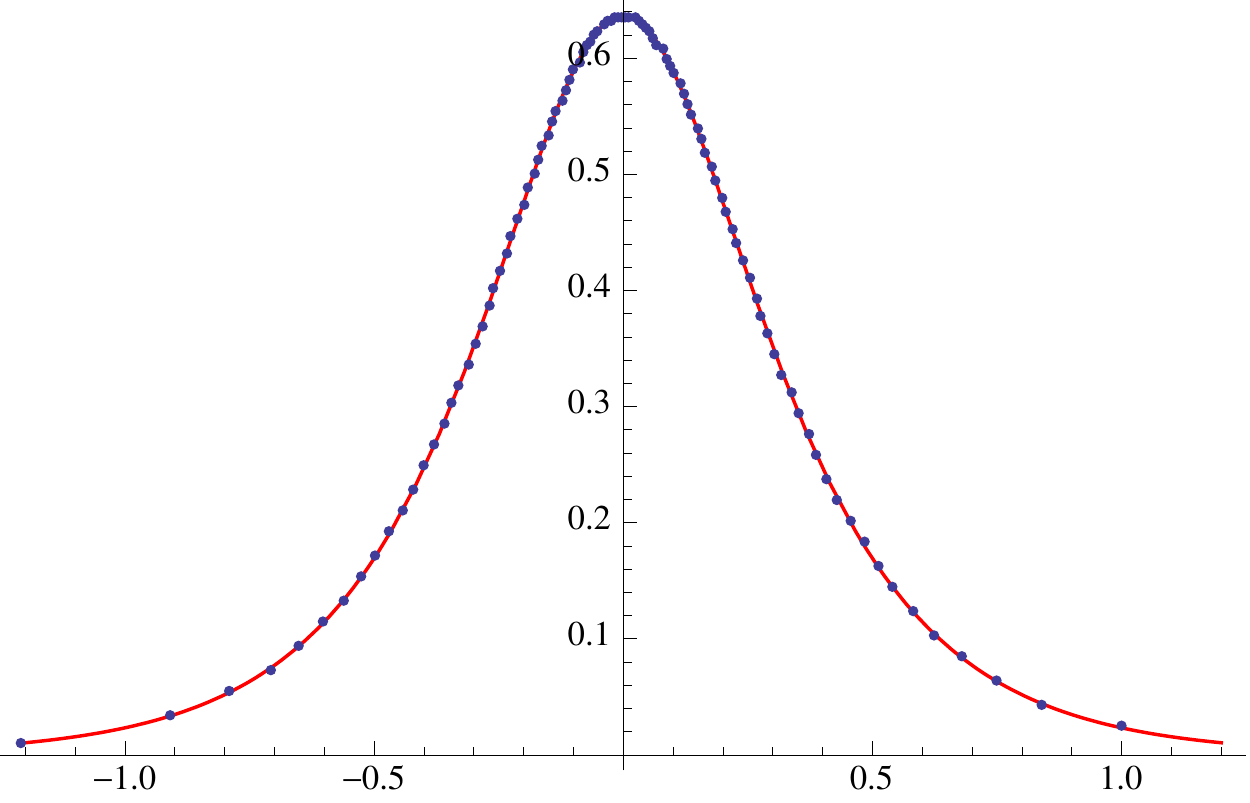}
  \caption{\textit{Left:} Bethe root distribution of the ground state with $\mathfrak t=-\i$ at $N=200$ sites. The roots condense along the lines $u = \pm 3\pi/8 +\i s$. \textit{Right:} Distribution of the Bethe roots along the line $u = 3\pi/8+\i s$. The points correspond to the numerical density, the solid line to the limiting distribution \eqref{eqn:densitybetheroots}.}
  \label{fig:rootdistribution}
\end{figure}

The remaining two cases $m=1,3$ are very different. First of all, we notice that with appropriate normalisation the solution for $m=3$ is simply the complex-conjugate of the solution for $m=1$. Hence we may restrict our considerations to $m=1$. A picture of the Bethe roots for $N=200$ sites is shown in figure \ref{fig:rootdistribution}. It indicates that they condense along the lines $u = \pm 3\pi/8 + \i s$.
The exact distribution along these lines can be derived from a differential equation. Indeed, the fact that $\mathcal R(u)$ is a sum of two hypergeometric functions implies that it solves a second-order ordinary differential equation. The equation takes its most convenient form if we introduce $r(u) = e^{\i(2-m)u}\mathcal R(u)/(\sin 4u)^{n-1/2}$ which solves
\begin{equation*}
  - r''(u) + \left(\frac{4(4n^2-1)}{\sin^2 4u}-\frac{1}{2}\left((4-m)^2+m^2\right)+\frac{\i}{2}\left((4-m)^2-m^2\right)\cot 4u\right)r(u)=0.
\end{equation*}
Setting now $g(s) = r(u=\pm 3\pi \i/8+\i s)$, we obtain for large $n$ to leading order the simple differential equation
\begin{equation*}
  g''(s) + \left(\frac{4n}{\cosh 4s}\right)^2 g(s)=0.
\end{equation*}
Suppose now that we find a certain real $s=s_0$ such that $g(s=s_0)=0$. In the close vicinity of $s_0$ the differential equation is similar to a classical harmonic oscillator with frequency $\omega = 2n/\cosh 4s_0$. It follows that the closest next zero can be found at a distance corresponding to a half-period $s_1 = s_0 + \pi/\omega$. If we thus introduce the Bethe root density $\rho_\pm(s_0) = f/(s_1-s_0)$, we obtain for large $f=2n$ the explicit distributions
\begin{equation}
  \rho_\pm(s) = \frac{2}{\pi \cosh 4s},
  \label{eqn:densitybetheroots}
\end{equation}
which agrees nicely with the numerical data (see figure \ref{fig:rootdistribution}).

\paragraph{Chains of length $\mathbf{N\neq 4n}$.} In this case we parametrise $N=4n+m$ with $m=1,2,3$. First of all, the constraints on $\mathfrak t$ lead to the following values
\begin{align*}
  \mathfrak t = (-1)^{m+1} ,\, \eta = (-1)^n2 \cos m\theta.
\end{align*}
As above we insert these values into the functional equation for $\mathcal R(u)$, and conclude that the only non-vanishing coefficients $c_j$ are those with $j=4k,j=4k+m$. Counting the remaining non-vanishing coefficients we find that there are $2(n+1)$ unknowns. As we need to impose that $\mathcal R(0) = \mathcal R'(0)=\cdots=\mathcal R^{(N-f-1)}(0)=0$, i.e. $N-f$ homogeneous linear equations, we obtain that a non-trivial solution is obtained for $f=2n+m-1$. Notice that these are precisely the fermion numbers of the ground states as determined in theorem \ref{thm:GS}. 
An explicit solution of the linear system up to a constant then leads to the solution
\begin{align*}
  \mathcal R(u) =& e^{-\i N u}\Biggl({_2}F_1\Bigl(-n,-n-\frac{m}{4},1-\frac{m}{4};e^{8\i u}\Bigr)\\
    &-\frac{\Gamma\left(1-\frac{m}{4}\right)}{\Gamma\left(1+\frac{m}{4}\right)} \frac{\Gamma\left(n+1+\frac{m}{4}\right)}{\Gamma\left(n+1-\frac{m}{4}\right)}e^{2\i m u}{_2}F_1\left(-n,-n+\frac{m}{4},1+\frac{m}{4};e^{8\i u}\right) \Biggr)  
\end{align*}
which is very similar to the cases at $N=4n$. For $m=2$ this reduces again to a very simple function $\mathcal R(u) = (-2\i \sin 2u)^{2n+1}$, and computing the $\mathcal Q$-function, we conclude again that the corresponding state has $f=2n+1$ coinciding Bethe roots at $u=2\theta =\pi/2$. We checked this for the first few cases explicitly, and thus extend our conjecture made above:
\begin{conjecture}
The ground state in the sector $f=N/2$ and $\mathfrak t=-1$ for $N=2,6,\dots$ for periodic boundary conditions is the limit of an appropriately renormalised Bethe state where all Bethe roots $u_1, u_2,\dots, u_f$ tend to $u=2\theta$.
\label{conj:condensation2}
\end{conjecture}
The other cases $m=1$ and $m=3$ are similar to the ones in the case where $N$ is a multiple of four. Their Bethe roots condense in the complex plane along the lines $u = \pm 3\pi/8 + \i s$ with real $s$, and the densities along these lines are given by \eqref{eqn:densitybetheroots}.

\subsubsection{Solution in the elliptic case}
The preceding considerations extend to the elliptic case where the chain is staggered, and $N=4n$ or $N=4n+2$. As the system is only invariant under translation by two sites, $\mathfrak t^2$ is the eigenvalue of the translation operator $T^2$ by two sites. Yet, we still have $\mathfrak t^N=1$ which implies that $\mathfrak t$ can take only discrete values. Hence, when deforming continuously the Hamiltonian from the translationally-invariant problem to the staggered case, the values of $\mathfrak t$ cannot change as they are discrete. We may therefore still characterise the three ground states for $N=4n$ by integers $m=1,2,3$ such that $\mathfrak t = \prod_{j=1}^{2n} z(u_j)^{-1} = -e^{2\i m \theta}$, whereas for $N=4n+2$ the single ground state has $\mathfrak t = -1$. Moreover, as the Hamiltonian conserves the number of fermions, we expect these ground states to contain $f=N/2$ particles.

Our aim is thus to solve the functional equation \eqref{eqn:recursionR} for $\mathcal R(u) =\Phi(u)\mathcal Q(u)$ in the elliptic case. There are two simple cases, namely $N=4n$ with $\mathfrak t=1$ and $N=4n+2$ with $\mathfrak t=-1$, for which the solution is given by
\begin{equation*}
  \mathcal R(u) = \Phi(u)\Phi(u-2\theta)= \left(\vartheta_1(u)\vartheta_1(u-2\theta)\right)^{N/2}.
\end{equation*}
It is therefore natural to extend the conjectures \ref{conj:condensation1} and \ref{conj:condensation2} about the condensation of the Bethe roots at a single point for these ground state to the elliptic case. The remaining two cases are more complicated. In order to solve the functional equation for $\mathcal R(u)$ we need to use its quasi-periodicity properties
\begin{equation*}
  \mathcal R(u+\pi) = (-1)^{N} \mathcal R(u), \quad \mathcal R(u+\pi \tau) = (-q^{-1})^N e^{-2\i Nu} e^{2\i \chi}\mathcal R(u),
\end{equation*}
where $\tau$ is related to the elliptic nome by $q=e^{\i \pi \tau}$, and we abbreviated the sum of all Bethe roots by $\chi = \sum_{j=1}^f u_j$. For the homogeneous model at $N=4n$, $\chi$ can be obtained directly from \eqref{eqn:RN4n}:
\begin{equation*}
  e^{2\i \chi(q=0)}= \frac{\sqrt{2}\pi}{\Gamma(1-m/4)^2}\frac{\Gamma(n+1-m/4)}{\Gamma(n+m/4)}
\end{equation*}
In the staggered case it is fixed through a highly non-trivial condition. To see this, we consider the Fourier expansion
\begin{equation*}
  \mathcal R(u) = e^{-\i N u}\sum_{j=-\infty}^\infty c_j e^{2\i j u}.
\end{equation*}
The functional equation for $\mathcal R(u)$ implies $c_j=0$ unless $j=4k, 4k+m$. Furthermore, the quasi-periodicity properties lead to $c_{j+N}=q^{-N+2 j}e^{-2\i \chi}c_j$, so that
\begin{equation*}
  c_{j+Nk} = e^{-2\i k \chi} q^{2j k +N k(k+1)/2} c_j.
\end{equation*}
This allows to rewrite
\begin{equation*}
  \mathcal R(u) = e^{-\i N u}\sum_{j=0}^{N-1} c_j e^{2\i j u}\Theta(Nu + 2j\pi \tau-\chi,q^N)
\end{equation*}
where we abbreviated $\Theta(u) = \sum_{j=-\infty}^\infty e^{2\i j u} q^{j(j-1)/2}$. For $N=4n$ the number of non-vanishing coefficients is $2n$. Because of $\mathcal R(0) = \mathcal R'(0)=\cdots=\mathcal R^{(N-f-1)}(0)=0$ with $f=2n$, they need to solve a homogeneous linear system of $2n$ equations whose coefficient matrix is a function of the unknown angle $\chi(q)$. For the linear system to admit a non-trivial solution the determinant of that matrix needs to vanish, which fixes $\chi(q)$ implicitly. The unknown coefficients can then be computed in terms of determinants, but do not seem to have a simple form even for small systems (this is similar to the observations made in \cite{fabricius:05}).

\section{Continuum theory, gap scaling and multiplets}
\label{sec:gap}

The purpose of this section is to argue that the staggered $M_2$ is a lattice version of the so-called super-sine-Gordon model. We support this by a discussion of its properties beyond the ground states by analysing the scaling function of its gap scaling, and furthermore the multiplet structure for excited states in various sectors of its Hilbert space. We restrict our considerations to periodic boundary conditions.

\subsection{Continuum theory of the model on the special submanifold}
In this section, we discuss the field theory interpretation of the special submanifold for which the model is integrable and has enhanced supersymmetry properties. In \cite{fendley:03,hagendorf:13} it was suggested that the continuum limit of the homogeneous model, i.e. the low energy effective theory in the limit of large chain length, is described by the second $\mathcal{N}=2$ superconformal minimal model. This superconformal field theory has a central charge, $c=3/2$, and can be understood as an Ising theory in terms of a free Majorana fermion, $\psi$, and a compact free boson, $\Phi \equiv \Phi+ 2 \pi r$, with compactification radius, $r=\sqrt{2}$ \cite{schwimmer:87,dixon:88}. Before we turn to the field theory interpretation of the special submanifold, let us explore some of the symmetry properties of the superconformal field theory. The operator content of this theory is given by the primary operators
\begin{equation*}
\sigma,\psi,
\end{equation*}
in the Ising sector with conformal weights $h_{\psi}=1/2$, $h_{\sigma}=1/16$, and similarly for the right-movers, $\bar{\sigma}$ and $\bar{\psi}$, and the vertex operators in the boson sector
\begin{equation*}
V_{m,n}=\, :e^{\i (m+n) \phi/r+\i (m-n)\bar{\phi}/r}:
\end{equation*}
with
\begin{equation*}
h_{m,n}=\frac{m+n}{2r^2}\quad  \text{and} \quad \bar{h}_{m,n}=\frac{m-n}{2r^2},
\end{equation*}
where $m, n \in \mathbb{Z}/2$ and we used the decomposition of the boson into left- and right-movers $\Phi=\phi+\bar{\phi}$. For $r=\sqrt{2}$ there are three supercharges in both the left- and right-moving sectors, therefore the theory is said to have $\mathcal{N}=(3,3)$ supersymmetry. The supercurrents read \cite{dixon:88}
\begin{align*}
& G^0=- \psi \partial \phi, \, \bar{G}^0= -\bar{\psi} \bar{\partial} \bar{\phi},\\
& G^{\pm} = \i \psi V_{\pm1,\pm1} = \i \psi :\exp(\pm \sqrt{2}\i \phi):,\\
& \bar{G}^{\pm} = \i \bar{\psi} V_{\pm1,\mp1}=\i \bar{\psi} :\exp(\pm \sqrt{2}\i \bar{\phi}):. 
\end{align*}
Note that this is consistent with the number of supercharges on the lattice, however, since the left- and right-moving sectors are not decoupled in the lattice model, the identification between lattice and continuum supercharges is not straightforward. Furthermore, it is clear that the ``spin-reversal'' symmetry of the lattice model is captured by the $\mathbb{Z}_2$ symmetry of the Ising sector of the theory. Finally, there are two conserved $U(1)$ currents, with charges $m$ and $n$, which are related to translation symmetry and particle number conservation on the lattice. Using the operator-state correspondence one can identify operators in the field theory with states in the lattice model. In particular, the three states with zero energy in periodic chains of length $N=4n$ are identified with the states
\begin{equation*}
\sigma \bar{\sigma} \ket{0}, \ V_{\pm1/2,0} \ket{0}.
\end{equation*}
where $|0\rangle$ denotes the conformal vacuum.
One easily verifies that these states have energy $E=h + \bar{h} -c/12=0$. From this identification one can also infer the relation between the lattice momentum, $p$, and particle number, $F$, on the one hand, and the $U(1)$ charges $m$ and $n$, on the other hand \cite{huijse:11_2}. We find
\begin{align*}
p &= \pi m + P + F \pi \mod 2 \pi,\\
F_A &= F-N/2 = n, 
\end{align*}
where $P=2\pi(h-\bar{h})/N$. Note that this is the momentum that appears in the supersymmetry algebra \ref{eqn:superalgebra}. Note that for $P=0$ we have $p=\pi (m + F) \mod 2\pi $. It follows that $T^4=\exp(4 p \i) = \exp(4 m \pi \i)=1$, where we used the fact that $m \in \mathbb{Z}/2$. We thus confirm that $P=0$ is consistent with $T^4 \equiv 1$ as anticipated in section \ref{sec:dynsusy}.

As discussed in the literature \cite{fendley:92,fendley:92_1,bajnok:04,hegedus:07}, this superconformal field theory has a relevant perturbation which is integrable and preserves supersymmetry. It is thus natural to identify the special submanifold in the vicinity of the homogeneous point, with the second $\mathcal{N}=2$ superconformal minimal model with this perturbation. In our notation the perturbing operator is given by the Neveu-Schwarz primary 
\begin{equation*}
\mathcal{O}_g = \psi \bar{\psi} (V_{1,0}+V_{-1,0}),
\end{equation*}
and has scaling dimensions $h=\bar{h}=3/4$. Here $g$ refers to the coupling parameter of this operator in the action. Note that this operator preserves the $\mathbb{Z}_2$ symmetry of the Ising sector, since it is basically a mass term for the Majorana fermion (however, with a mass depending on $\Phi$). Furthermore, this operator has momentum $p=\pi$ and therefore the action is no longer invariant under the action of the translation operator $T=\exp(\i p)$, it is, however, invariant under the action of $T^2$. Consequently, these symmetry properties nicely agree with those of the special submanifold away from the homogeneous point. Finally, one may also verify that this perturbation indeed preserves all the supercurrents to first order in $g$ in perturbation theory \cite{bernard:90_2}. There are no other relevant operators that also preserve all these symmetries as required by our model. We thus conclude that moving away from the homogeneous point on the special submanifold indeed corresponds with turning on a coupling to this operator. The renormalisation group flow induced by this relevant coupling takes us to the super-Sine-Gordon model \cite{bajnok:04}. It follows that the continuum theory which describes the model on the special submanifold for $N\to \infty$ in an appropriate scaling window around the homogeneous point is the super-Sine-Gordon theory.

\subsection{Gap scaling and dynamical supersymmetry}

In this section, we show that the presence of the dynamical supersymmetry allows us to infer the leading behaviour of the gap scaling function. We start with some general arguments on gap scaling close to a critical point. Suppose that we perturb the fixed point action with an operator $\mathcal{O}_g$ with scaling dimension $\Delta_g =h_g+ \bar{h}_g$ and coupling parameter $g$. We then find that close to the critical point the energy relative to the ground state, $E_N$, of the system of size $N$ behaves as
\begin{equation}
E_N(g)-E_N(0)=f(z)/N + \dots,\label{eq:scaling}
\end{equation}
where $z$ is a dimensionless parameter that depends on the coupling: $z=g N^y$, with $y\equiv 2- \Delta_g$. The function $f$ is the gap scaling function and the dots denote subleading terms that vanish as $N \to \infty$. 

\begin{figure}[h]
\centering
\includegraphics[width=.7\textwidth]{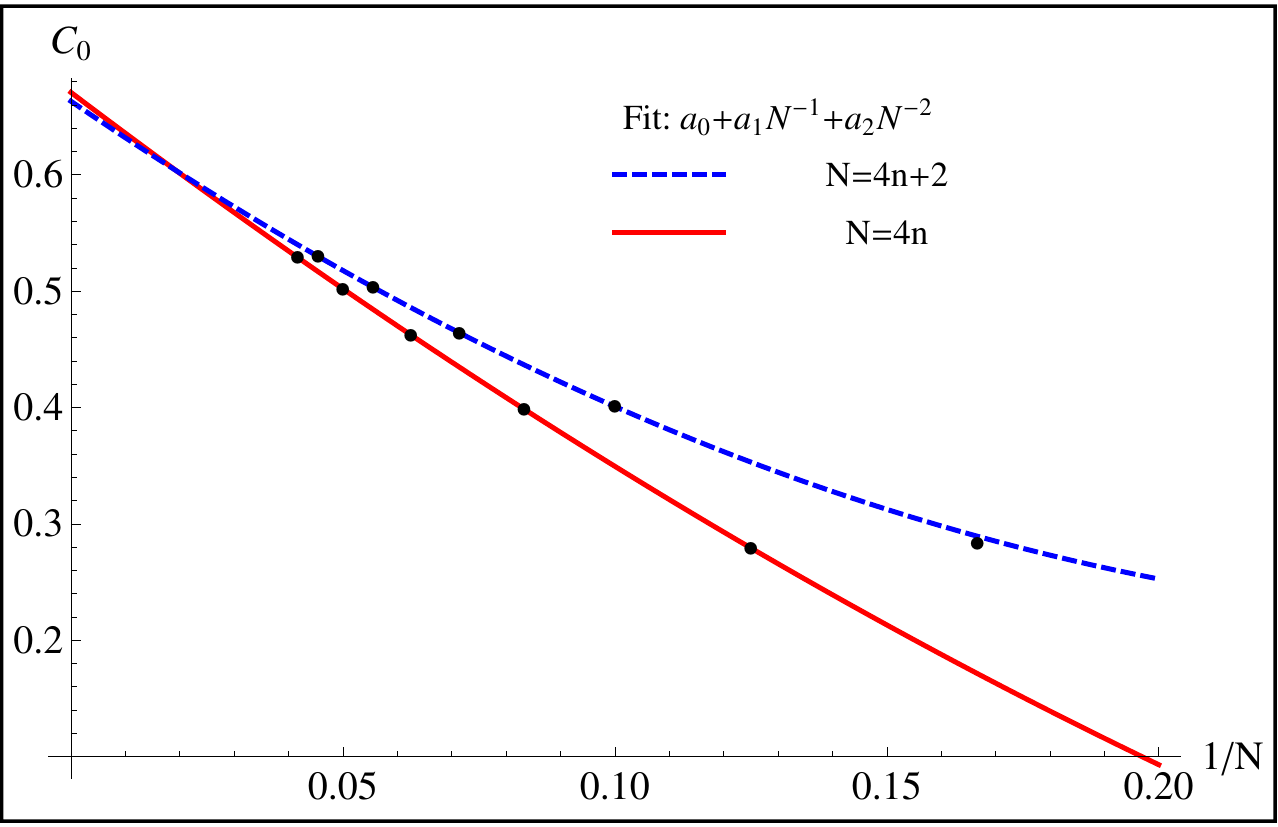}
\caption{We plot the fit parameters $C_0$ versus $1/N$. The data is fitted to $a_0 + a_1 N^{-1} + a_2 N^{-2}$ for $N=4n$ and $N=4n+2$ separately. In both cases we find $a_0 \approx 0.66$.  \label{fig:fitscaling}}
\end{figure}

Now consider a lattice Hamiltonian with a dynamical supersymmetry that changes the length of the system by $n$ sites. It follows that all the states of a system with $N$ sites, have superpartners in the system with either $N-n$ or $N+n$ sites. These superpartners have the same energy. That is, we have
\begin{equation*}
E_N(g)-E_N(0)=E_{N+n}(g)-E_{N+n}(0).
\end{equation*}
At the same time, the scaling form (\ref{eq:scaling}) holds for both energy levels. Note that, since the levels are degenerate, the scaling function, $f$, has to be identical for both levels. However, the corrections to the asymptotic form can be different for the two levels. We thus find,
\begin{align*}
\frac{f(z)}{N}&=\frac{f(g(N+n)^y)}{N+n}
\end{align*}
Expanding this equation for large $N$ with fixed scaling variable $z=gN^y$ we obtain
\begin{align*}
\left(1+\frac{n}{N}\right)f(z)&= f(g (N+n)^{y}) =f\left(g N^y \left(1+\frac{ny}{N}+\dots\right)\right)=f(z)+\frac{n}{N}zy f'(z) +\dots, \nonumber
\end{align*}
where the dots correspond to subleading corrections.
To leading order we thus find
\begin{equation}
f(z) = zy f'(z) \Rightarrow f(z) = a_0 z^{1/y},  \label{eq:diffrel}
\end{equation}
with $a_0$ an unknown constant. The dynamical supersymmetry thus imposes a non-trivial constraint on the shape of the scaling function. In particular, for our model we have $\Delta_g=3/2$, which implies $y=1/2$. 
\begin{figure}[h]
\centering
\includegraphics[width=.8\textwidth]{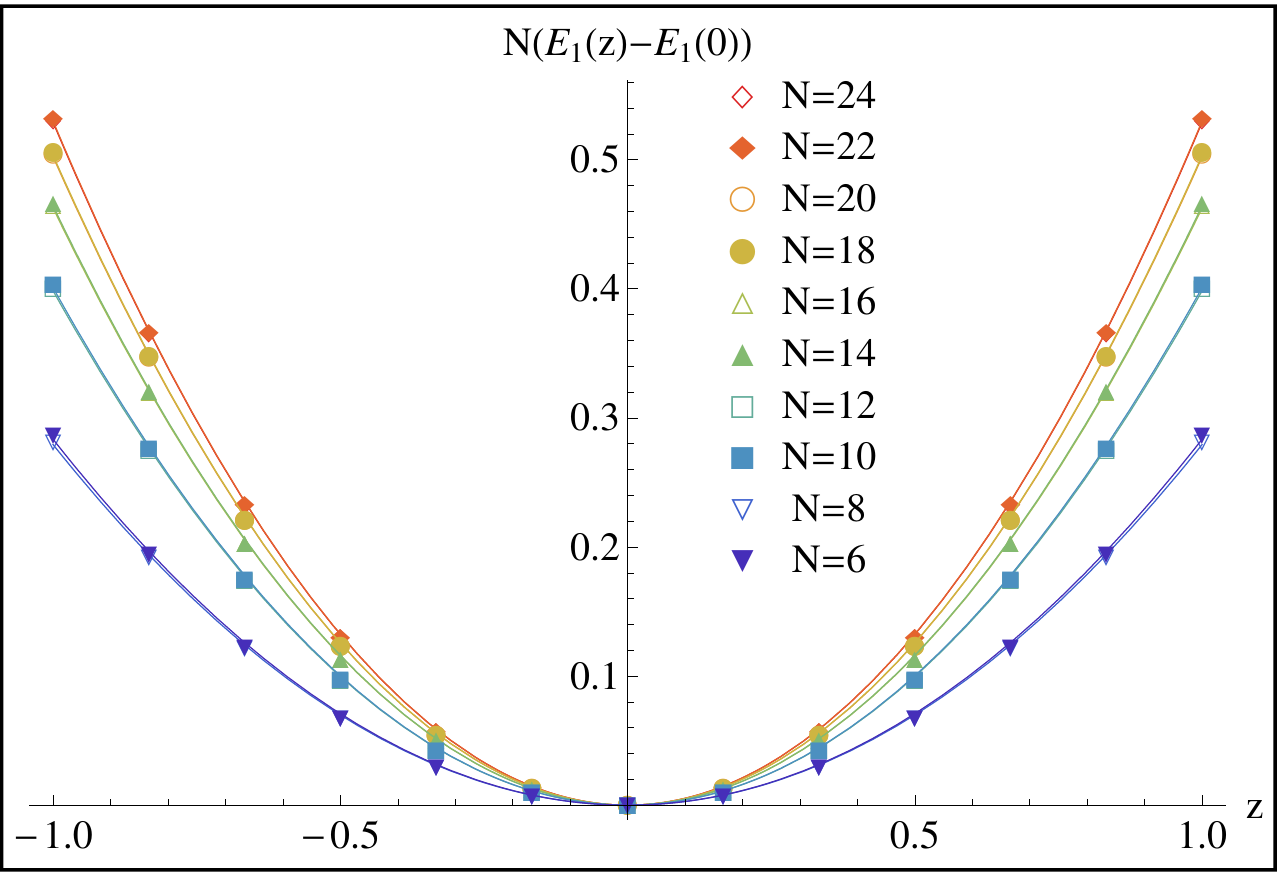}
\caption{We plot $N(E_1 (g)- E_1(0))$ versus $z=g N^y$ for $L=6,8,\dots,24$, where $E_1$ is the energy of the first excited state, the coupling $g \equiv 2 \mu^2-1$ is zero at the homogeneous point and $y=1/2$. The lines are parabolic fits to the function $C_0 z^2$ with $C_0$ the fit parameter. \label{fig:scaling}}
\end{figure}
In figure \ref{fig:scaling} we have plotted the scaling function for the energy of the first excited state of systems of even length, $N$, in the sector with fermion number $F=N/2$ and translation eigenvalue $\mathfrak t^2=1$. We restrict to the special submanifold and, in particular, we take $\lambda_x=1, \ \mu_x=\mu_{x+2} \ \forall x$ and $\mu_1=1, \ \mu_2=\mu$. We plot $N(E_1 (g)- E_1(0))$ versus $z=g N^y$, where $E_1$ is the energy of the first excited state, the coupling $g \equiv 2 \mu^2-1$ is zero at the homogeneous point and $y=1/2$. Since $f(z)=N(E_1 (g)- E_1(0))+h(z,N)$, where $h(z,N)$ is a finite-size correction that tends to zero as $N \to \infty$, we expect the data plotted in this way to collapse on one curve for large enough system sizes. In particular, as we just derived, we expect this curve to be quadratic in $z$. It is clear that the system sizes are too small to see data collapse. However, the quadratic dependence on $z$ is clear; the lines are fits to the function $C_0 z^2$ with $C_0$ a fit parameter. This fit parameter depends on the length, in fact, we expect
\beq
C_0=z^{-2}(f(z)-h(z,N)) = a_0 + a_1 N^{-1} + a_2 N^{-2}+ \mathcal{O}(N^{-3}). \nonumber
\eeq
In figure \ref{fig:fitscaling} we have plotted the fit parameters $C_0$ versus $1/N$. We see a clear even-odd effect and when we fit the data for $N=4n$ and $N=4n+2$ to $a_0 + a_1 N^{-1} + a_2 N^{-2}$, we find $a_0 \approx 0.66$ in both cases. This is a very good indication that for large enough system sizes, the gap scaling data will indeed collapse onto the curve $f(z)=a_0 z^2$ with $a_0  \approx 0.66$.

\subsection{Multiplet Structure}
The eigenstates of the Hamiltonian organise in representations of its symmetry algebra. From our discussion in section \ref{sec:susy}, we conclude that the amount of supersymmetry of the model on the special submanifold actually depends on the translation subsector of the Hilbert space we consider. Here we provide a brief analysis of the resulting multiplet structure in the case of periodic boundary conditions.

First of all, by definition the model always contains a copy of the $\mathcal N=2$ supersymmetry algebra with non-dynamic supercharges $Q_+,\bar Q_+$. As the Hamiltonian is given by their anticommutator, and thus a positive operator, its eigenvalues are either zero or positive. It is well-known that the zero-energy eigenstates correspond to supersymmetry singlets, i.e. eigenstates which are annihilated by both supercharges. Conversely, all positive eigenvalues are doubly-degenerate, and the corresponding eigenvalues come as doublets $(|\psi\rangle,\,Q_+|\psi\rangle)$ where $\bar Q_+|\psi\rangle =0$.

Second, in translation sectors where $T^4\equiv 1$, the multiplets have a richer structure because of the presence of a second $\mathcal N=2$ algebra with dynamic supercharges $Q_-,\bar Q_-$. Since $[T^2,H]=0$, we need to distinguish two cases here. \textit{(i)} If $T^2 \equiv -1$, which can be the case only if the number of sites is a multiple of four, we find that states with positive energy organise in quartets $(|\psi\rangle, Q_+|\psi\rangle, Q_-|\psi\rangle,Q_+Q_-|\psi\rangle)$ where $|\psi\rangle$ is annihilated by both $\bar Q_+$ and $\bar Q_-$. So-called short (or BPS) multiplets are not present, because the central charges of the algebra generated by the supercharges are all zero. \textit{(ii)} If on the other hand $T^2\equiv 1$, then a third dynamic copy of the supersymmetry algebra generated by $Q_0, \bar Q_0$ is present. In this case, the non-zero energy states organise in octets, which are generated from a cyclic state, that is, a state annihilated by all the adjoint supercharges $\bar Q_+,\,\bar Q_-,\,\bar Q_0$. 

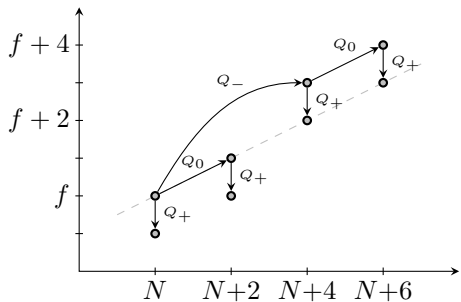
\begin{figure}[h]
  \centering
  \begin{tikzpicture}[>=stealth,
  place/.style={circle,draw=black,fill=lightgray,thick,
                 inner sep=0pt,minimum size=1mm},
   transition/.style={rectangle,draw=black!50,fill=black!20,thick,
                      inner sep=0pt,minimum size=4mm}]
  \draw[lightgray,dashed] (-.5,0.25)--(3.5,2.25);
                      
  \node (cyclic) at ( 0,.5) [place] {};
  \node (Qcyclic) at ( 0,0) [place] {};
  \node (Q0cyclic) at ( 1,1) [place] {};
  \node (QQ0cyclic) at ( 1,.5) [place] {};
  \node (Qbarcyclic) at ( 2,2) [place] {};
  \node (QQbarcyclic) at ( 2,1.5) [place] {};
  \node (Q0Qbarcyclic) at ( 3,2.5) [place] {};
  \node (QQ0Qbarcyclic) at ( 3,2) [place] {};
  
  \draw[->] (cyclic) -- (Q0cyclic);
  \draw[->] (cyclic) -- (Qcyclic);
  \draw[->] (Q0cyclic) -- (QQ0cyclic);
  \draw[->] (Qbarcyclic) -- (Q0Qbarcyclic);
  \draw[->] (Qbarcyclic) -- (QQbarcyclic);
  \draw[->] (Q0Qbarcyclic) -- (QQ0Qbarcyclic);
  
  \draw[->] (cyclic) to [out=60, in = 180] (Qbarcyclic);
  
  \draw[<->] (-1,3) -- (-1,-.5) -- (4,-.5);
  \foreach \y in {0,.5,...,2.5}
    \draw (-1.05,\y) -- (-.95,\y);
  \foreach \x in {0,1,2,3}
    {
    \draw (\x,-.55) -- (\x,-.45);
    }
    \draw (0,-.5) node[below] {$N$};
    \draw (1,-.5) node[below] {$N{+}2$};
    \draw (2,-.5) node[below] {$N{+}4$};
    \draw (3,-.5) node[below] {$N{+}6$};
    \draw (-1,2.5) node[left] {$f+4$};
    \draw (-1,1.5) node[left] {$f+2$};
    \draw (-1,.5) node[left] {$f$};
    
    \draw(1,2) node {\tiny $Q_-$};
    \draw (2.5,2.5) node {\tiny $Q_0$};
 \draw (0.5,.95) node {\tiny $Q_0$};
 \draw (.3,.25) node {\tiny $Q_+$}; 
\draw (1.3,.75) node {\tiny $Q_+$}; 
\draw (2.3,1.75) node {\tiny $Q_+$}; 
\draw (3.3,2.25) node {\tiny $Q_+$};

\end{tikzpicture}
  \caption {Structure of a multiplet and its quantum numbers at positive energy: all states are generated through the action of the supercharges on the cyclic state. The dashed line corresponds to states with the same $F_A$ as the cyclic state of the multiplet.\label{fig:multiplet}}
\end{figure}

\paragraph{Multiplet structure in numerical spectra.}
The multiplet structure can be observed directly in spectra obtained from exact diagonalisation of the Hamiltonian. In this section we present the numerical data for three different choices of the parameters $\lambda_x$ and $\mu_x$, corresponding to the homogeneous point on the special submanifold (red dot in Fig. \ref{fig:param}), a different point on the special submanifold (blue dot  in Fig. \ref{fig:param}), and finally, for comparison a point away from the special submanifold (green dot in Fig. \ref{fig:param}).

\begin{figure}[h]
\includegraphics[width=.95\textwidth]{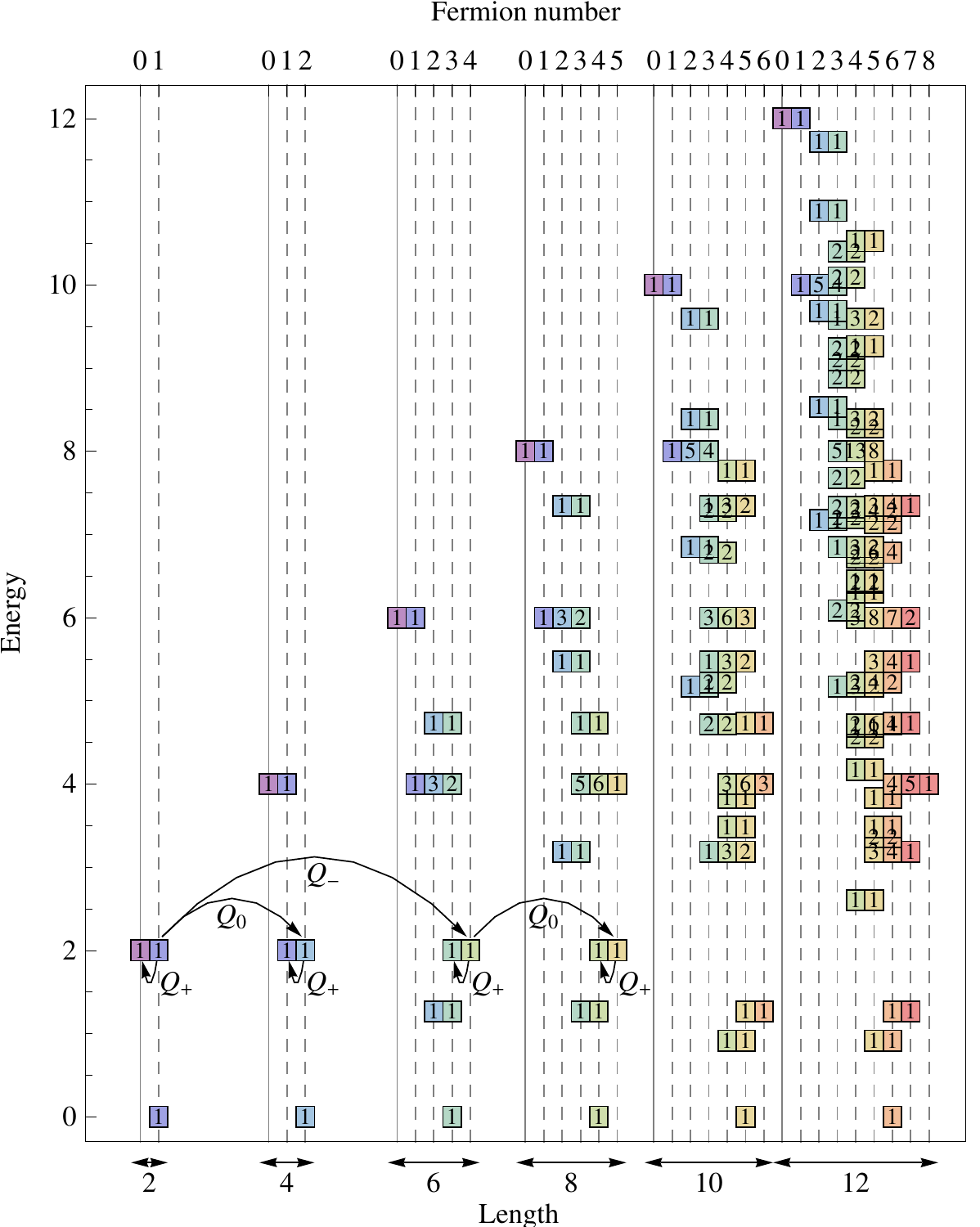}
\caption{The plot shows the energy levels of chains of even lengths $N=2,4,\dots,12$ for the translation sector $\mathfrak t^2=1$. The parameters in the Hamiltonian are taken at the homogeneous point, i.e. $\lambda_x=1, \mu_x=1/\sqrt{2}\ \forall \ x$. To reveal the multiplet structure we split up the spectrum of a single chain length by fermion number. At the bottom of the plot we indicate a range containing the levels of a chain of a certain length. At the top of the plot we indicate the fermion number, the grey dashed and drawn lines serve as a guide to the eye, where the drawn lines correspond to fermion number $f=0$. The vertical axis labels the energy. We plot each level as a box, where the numerical value of the energy corresponds to the centre of the box. The numbers inside the boxes give the degeneracy of the level and the colours of the boxes are one-to-one with the fermion numbers. \label{fig:spectraTrig}}
\end{figure}

In figure \ref{fig:spectraTrig}, we plot numerical data obtained by exact diagonalisation of the Hamiltonian. In particular, we plot the energy levels of chains of even lengths $N=2,4,\dots,12$ for the translation sector $\mathfrak t^2=1$. The parameters in the Hamiltonian are taken at the homogeneous point, i.e. $\lambda_x=1, \mu_x=1/\sqrt{2}\ \forall \ x$. To reveal the multiplet structure we split up the spectrum of a single chain length by fermion number. The symmetry generated by $Q_+$ is very obvious in this plot, but also the full multiplets, octets in this case, can be detected upon closer examination. For one of the multiplets, with the cyclic state at $N=2, f=1, E=2$, we indicate the action of the supercharges. This can be compared directly with figure \ref{fig:multiplet}. The multiplets that start at length $N=8$ or larger are incomplete, because we truncate the spectrum at length $N=12$.

\begin{figure}[h]
\includegraphics[width=.95\textwidth]{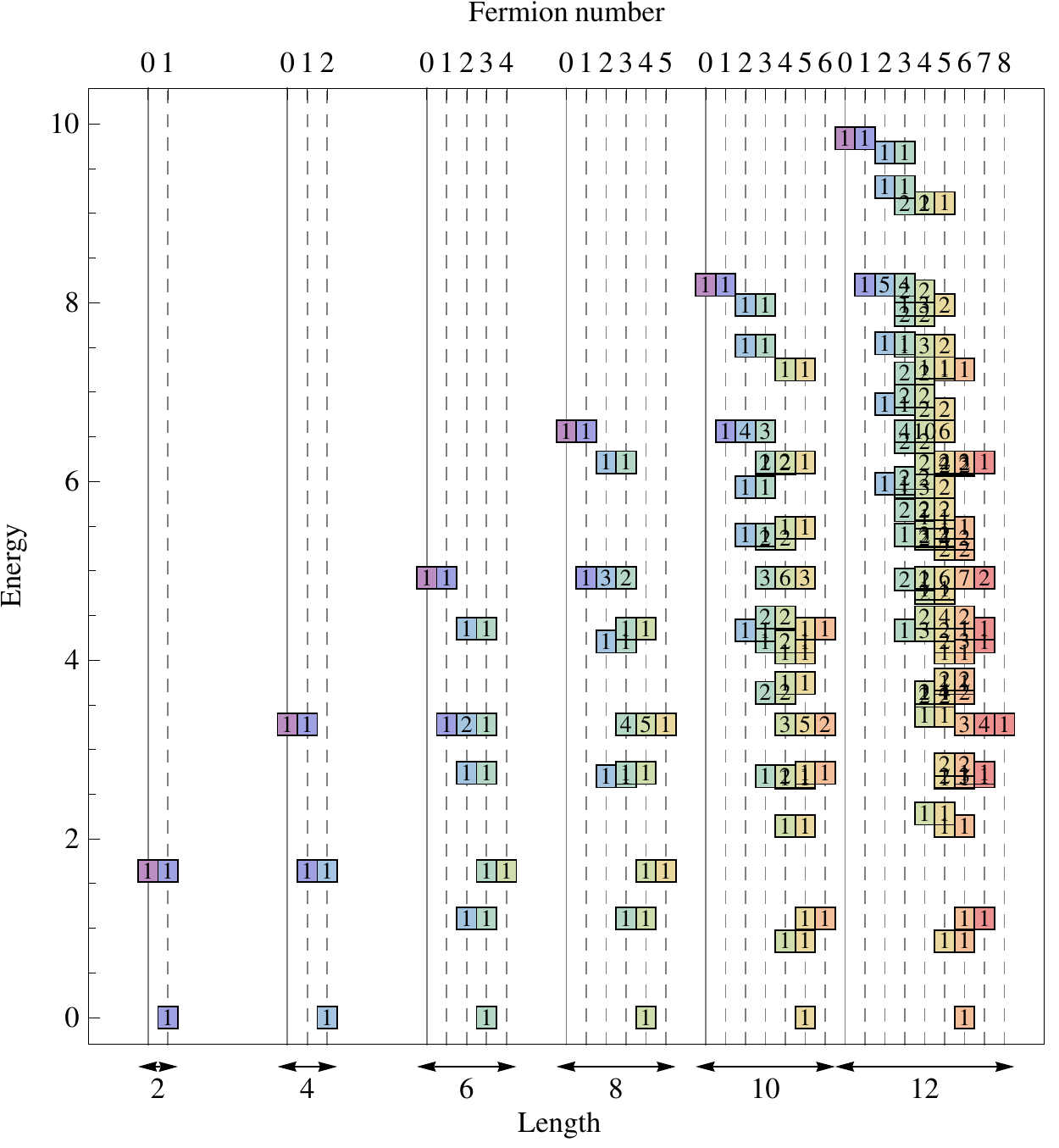}
\caption{The plot shows the energy levels of chains of even lengths $N=2,4,\dots,12$ for the translation sector $\mathfrak t^2=1$. The parameters in the Hamiltonian are taken at some point on the manifold with dynamical supersymmetry, specifically $\lambda_x=4/5 \lambda_{x+1}=1$, $\mu_x=1/2$ and $\mu_{x+1}=\sqrt{3}/2$ for all even $x$. See figure \ref{fig:spectraTrig} for details on how to read the plot. \label{fig:spectraCyl}}
\end{figure}

\begin{figure}[h]
\includegraphics[width=.95\textwidth]{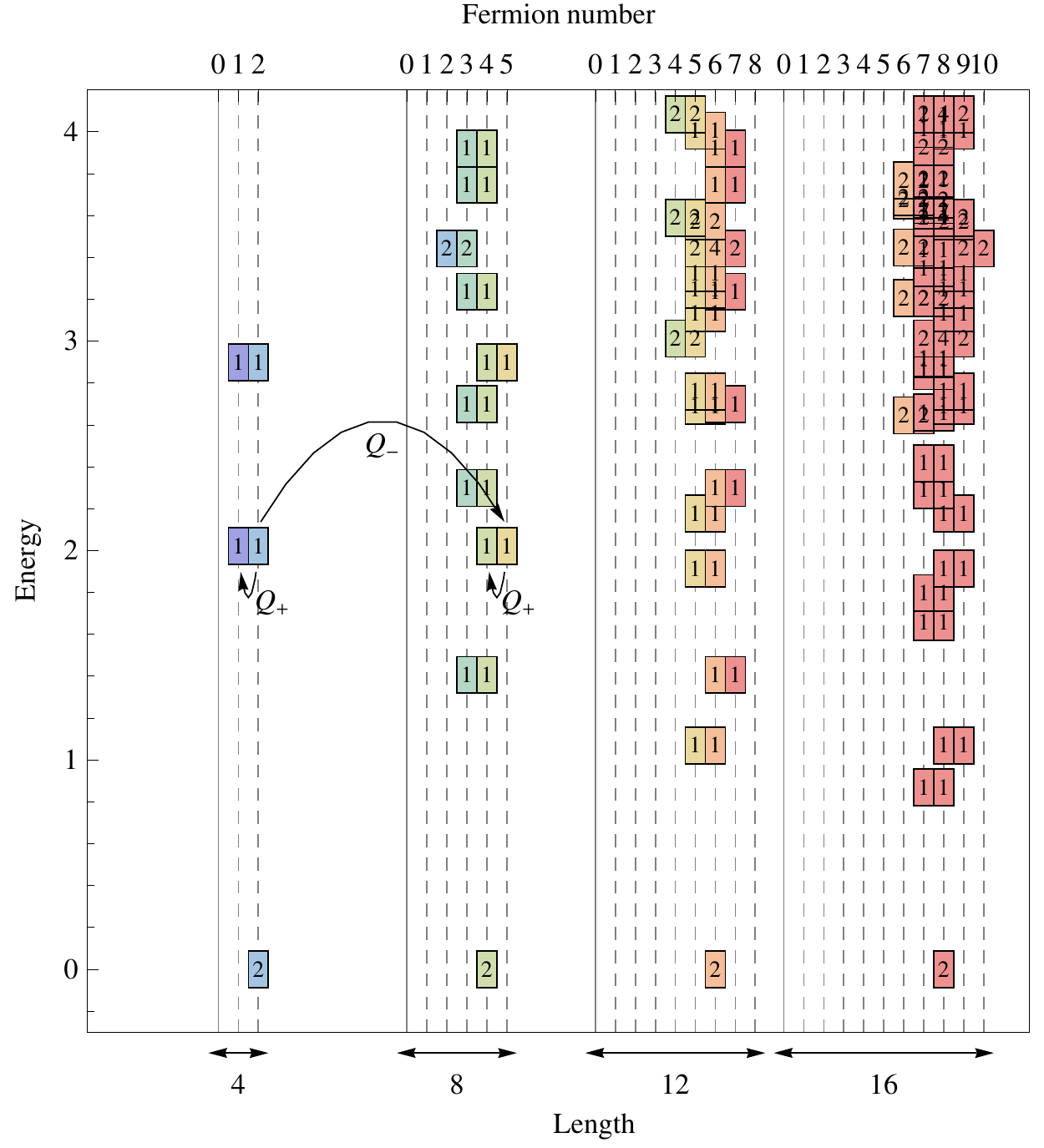}
\caption{The plot shows the energy levels of chains of even lengths $N=4,8,\dots,16$ for the translation sector $\mathfrak t^2=-1$. The parameters in the Hamiltonian are taken at some point on the manifold with dynamical supersymmetry, specifically $\lambda_x=4/5 \lambda_{x+1}=1$, $\mu_x=1/2$ and $\mu_{x+1}=\sqrt{3}/2$ for all even $x$. The spectrum is cut-off at $E=4.1$. See figure \ref{fig:spectraTrig} for details on how to read the plot. \label{fig:spectraCylTmin}}
\end{figure}

\begin{figure}[h]
\includegraphics[width=.95\textwidth]{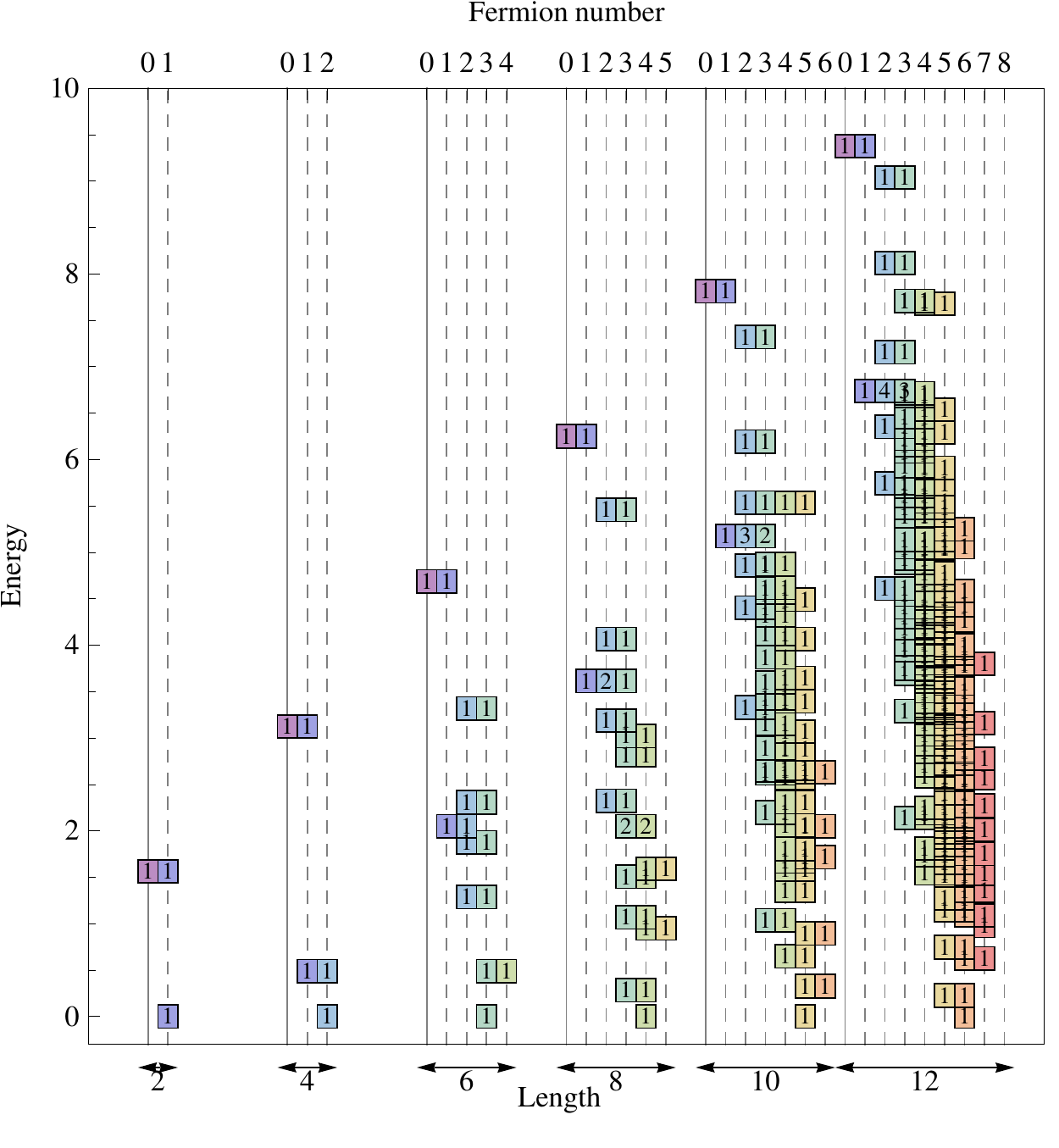}
\caption{The plot shows the energy levels of chains of even lengths $N=2,4,\dots,12$ for the translation sector $\mathfrak t^2=1$. The parameters in the Hamiltonian are staggered with period 2, but do not lie on the manifold with dynamical supersymmetry, specifically $\lambda_x=3/4 \lambda_{x+1}=1$, $\mu_x=1/4$ and $\mu_{x+1}=1/2$ for all even $x$. See figure \ref{fig:spectraTrig} for details on how to read the plot. \label{fig:spectraOff}}
\end{figure}

To show that the multiplet structure survives when we move away from the homogeneous point, while remaining on the manifold with dynamical supersymmetry, we show the same plot for $\lambda_x=4/5 \lambda_{x+1}=1$, $\mu_x=1/2$ and $\mu_{x+1}=\sqrt{3}/2$ for all even $x$ for the translation sectors $\mathfrak t^2=1$ and $\mathfrak t^2=-1$ (see Figs. \ref{fig:spectraCyl} and \ref{fig:spectraCylTmin}). The supermultiplets in the translation sector  $\mathfrak t^2=-1$ consist of at most 4 elements, since $Q_0$ is absent in this sector. Finally, in figure \ref{fig:spectraOff} we plot the spectra for a choice of parameters away from the manifold with dynamical supersymmetry. The parameters are staggered with period 2, since this is required in order for $T^2$ to commute with the Hamiltonian. It is clear that the doublet structure generated by $Q_+$ is preserved, while the multiplets corresponding to the dynamical supersymmetry are  absent.

Finally, we point out that the supersymmetries still do not account for all the degeneracies in the spectrum (see for instance the degeneracies of the levels with $E=4$ in figure \ref{fig:spectraTrig}). This suggests the possible presence of further symmetries.

\section{Conclusion}
\label{sec:conclusion}

In this work we provided a detailed analysis of the one-dimensional $M_2$ model of strongly-interacting fermions and pairs. In particular, we determined a submanifold in the space of parameters for which the model presents two hidden dynamic supersymmetries. We showed that this symmetry enhancement is present precisely for the choice of parameters for which the model is also diagonalisable by the Bethe ansatz. This allowed to understand its various symmetries in terms of the Bethe equations: in particular, we pointed out a relation between the existence of a dynamic supersymmetry and the existence of exact strings. Moreover, we determined the number of ground states for finite systems through cohomology techniques and found the distribution of their Bethe roots. Finally, we argued that the model provides a lattice version of the super-sine-Gordon model and obtained its gap scaling function from pure symmetry considerations.

It will be interesting to analyse the Bethe ansatz presented here in more detail in order to analyse certain physical properties of the model. Let us give a few examples of open problems. First, the study of spectral flow for the $M_2$ model through variations of the twist angle appears to be a natural extension, in particular the understanding of how the ground states flow, how the Ramond sector (periodic boundary conditions) and Neveu-Schwarz sector (antiperiodic boundary conditions) are related on the lattice, and how their Bethe roots evolve. We expect this to give new insights into how (superconformal) field-theory concepts \cite{lerche:89} are already realised in the lattice model. Second, we noted that the Bethe ansatz for the staggered $M_2$ model suggests that the model possesses an isospectral symmetry. It would be interesting to understand this symmetry, and its impact on the calculation of simple correlation functions such as sublattice occupation densities. Moreover, certain correlation functions are expected to have the scale-free property which was first observed in the related $M_1$ model \cite{fendley:10}. It remains still to understand its relation to the Bethe ansatz. Third, we argued how the scaling function for the energy gap off the critical point can be obtained from symmetry considerations. It would be interesting to investigate this point directly through the Bethe ansatz.

Furthermore, there are a few aspects beyond the $M_2$ model. One of the key features to detect the Bethe-ansatz solvability was an asymptotic analysis for small and large rapidities within the one- and two-particle problem of the Bethe ansatz. This strategy has potential to be applied to other systems, and help find integrable points when only a Hamiltonian (without additional structures such as transfer matrices etc.) is given. In fact, the reasoning presented here can be extended to all $M_k$ models with $k=1,2,3,\dots$ This problem will be addressed in a future publication \cite{chlh:tbp}. Eventually, the relation between dynamical symmetries and exact strings which we encountered in the model studied in this article deserves a deeper, and more general investigation. The recent progress to classify integrable (super)spin chains with dynamic supersymmetry was built on the idea that this type of symmetry presents itself through certain degeneracies in the Bethe equations at certain root-of-unity points \cite{meidinger:13}. Exact strings of Bethe roots play an important role at these points, and it will be interesting to understand if they allow the construction of larger supersymmetry algebras.

\subsection*{Acknowledgements} CH acknowledges support from the Belgian Interuniversity Attraction Poles Program P7/18 through the network DYGEST (Dynamics, Geometry and Statistical Physics), and would like to thank the Stanford Institute for Theoretical physics, where part of this work was done, for hospitality. TBF is supported by the Netherlands Organisation for Scientific Research (NWO). LH is funded by the John Templeton Foundation and a DOE early career award. We would like to thank Bernard Nienhuis for discussions.


\begin{thebibliography}{10}

\bibitem{fendley:03}
P.~{Fendley}, B.~{Nienhuis}  and K.~{Schoutens},
\newblock {\em {Lattice fermion models with supersymmetry}},
\newblock J. Phys. A: Math. Gen. {\textbf{36}} (2003)   12399--12424,
\newblock arXiv:cond-mat/0307338.

\bibitem{VW}
G.~Veneziano and J.~Wosiek,
\newblock {\em A supersymmetric matrix model: {III.} hidden susy in statistical
  systems},
\newblock JHEP {\textbf{2006}} {\textbf{11}} (2006)   030.

\bibitem{grover:14}
T.~Grover, D.~N.~Sheng  and A.~Vishwanath,
\newblock {\em Emergent space-time supersymmetry at the boundary of a
  topological phase},
\newblock Science {\textbf{344}} (2014)   280--283.

\bibitem{HuijseBauerBerg14}
L.~{Huijse}, B.~{Bauer}  and E.~{Berg},
\newblock {Emergent supersymmetry at the Ising-Berezinskii-Kosterlitz-Thouless
  multicritical point},
\newblock arxiv: 1403.5565 2014.

\bibitem{fendley:03_2}
P.~{Fendley}, K.~{Schoutens}  and J.~{de Boer},
\newblock {\em {Lattice Models with N=2 Supersymmetry}},
\newblock Phys. Rev. Lett. {\textbf{90}} (2003)   120402,
\newblock arXiv:hep-th/0210161.

\bibitem{fendley:05_2}
P.~{Fendley} and K.~{Schoutens},
\newblock {\em {Exact Results for Strongly Correlated Fermions in $2{+}1$
  Dimensions}},
\newblock Phys. Rev. Lett. {\textbf{95}} {\textbf{4}} (July 2005)   046403,
\newblock arXiv:cond-mat/0504595.

\bibitem{HvE}
H.~van Eerten,
\newblock {\em Extensive ground state entropy in supersymmetric lattice
  models},
\newblock J. Math. Phys. {\textbf{46}} (2005)   123302.

\bibitem{JJ}
J.~Jonsson,
\newblock {\em Certain homology cycles of the independence complex of grids},
\newblock Discrete Comput. Geom. {\textbf{43}} (2010)   927--950.

\bibitem{Tri12}
L.~Huijse, D.~Mehta, N.~Moran, K.~Schoutens  and J.~Vala,
\newblock {\em Supersymmetric lattice fermions on the triangular lattice:
  superfrustration and criticality},
\newblock New J. Phys {\textbf{14}} (2012)   073002.

\bibitem{hagendorf:13}
C.~Hagendorf,
\newblock {\em {Spin chains with dynamical lattice supersymmetry}},
\newblock J. Stat. Phys. {\textbf{150}} (2013)   609--657.

\bibitem{meidinger:13}
D.~{Meidinger} and V.~{Mitev},
\newblock {Dynamic Lattice Supersymmetry in gl(n|m) Spin Chains},
\newblock arXiv:1312.7021 2013.

\bibitem{baxterbook}
R.J.~Baxter,
\newblock {\em {Exactly solved models in statistical mechanics}},
\newblock London Academic, 1982.

\bibitem{blom:12}
L.~C. Blom,
\newblock {\em {Supersymmetry on a chain: A handle on strongly interacting
  fermions}},
\newblock Master's thesis, Unversiteit van Amsterdam, 2012.

\bibitem{fendley:10}
P.~{Fendley} and C.~{Hagendorf},
\newblock {\em {Exact and simple results for the XYZ and strongly interacting
  fermion chains}},
\newblock J. Phys. A: Math. Theor. {\textbf{43}} (2010)   402004.

\bibitem{zamolodchikov:81}
A.~B. Zamolodchikov and V.~A. Fateev,
\newblock {\em {A model factorized $S$-matrix and an integrable spin-$1$
  Heisenberg chain}},
\newblock Sov. J. Nucl. Phys. {\textbf{32}} (1981)   298--303.

\bibitem{fateev:81}
V.~A. Fateev,
\newblock {\em {A factorized $S$-matrix for particles of opposite parities and
  an integrable $21$-vertex statistical model}},
\newblock Sov. J. Nucl. Phys. {\textbf{33}} (1981)   761--766.

\bibitem{hori:03}
K.~Hori, S.~Katz, A.~Klemm, R.~Pandharipande, R.~Thomas, C.~Vafa, R.~Vakil and E.~Zaslow,
\newblock {\em {Mirror symmetry}},
\newblock Amer. Math. Soc., 2003.

\bibitem{crampe:11}
N.~Cramp\'e, E.~Ragoucy  and L.~Alonzi,
\newblock {\em {Coordinate Bethe ansatz for spin $s$ XXX model}},
\newblock SIGMA {\textbf{7}} (2011)  ~6.

\bibitem{whittaker:27}
E.~T. Whittaker and G.~N. Watson,
\newblock {\em A course of modern analysis},
\newblock Cambridge University Press, 1927.

\bibitem{chlh:tbp}
C.~Hagendorf and L.~Huijse.
\newblock {Bethe ansatz for the $M_k$ models of strongly-interacting
  fermions with supersymmetry}.
\newblock Manuscript (2014).

\bibitem{baxter:02}
R.~J. Baxter,
\newblock {\em {Completeness of the Bethe Ansatz for the Six and Eight-Vertex
  Models}},
\newblock J. Stat. Phys. {\textbf{108}} (2002)   1--48.

\bibitem{witten:82}
E.~Witten,
\newblock {\em {Constraints on supersymmetry breaking}},
\newblock Nucl. Phys. B {\textbf{202}} (1982)   253 -- 316.

\bibitem{beccaria:12}
M.~Beccaria and C.~Hagendorf,
\newblock A staggered fermion chain with supersymmetry on open intervals,
\newblock arXiv:1206.4194 2012.

\bibitem{huijse:11}
L.~{Huijse}, N.~{Moran}, J.~{Vala}  and K.~{Schoutens},
\newblock {\em {Exact ground states of a staggered supersymmetric model for
  lattice fermions}},
\newblock Phys. Rev. B {\textbf{84}} (2011)   115124,
\newblock arXiv:1103.1368.

\bibitem{huijse:10}
L.~Huijse,
\newblock {\em A supersymmetric model for lattice fermions},
\newblock PhD thesis, Universiteit van Amsterdam, 2010.

\bibitem{BottTu82}
R.~Bott and L.W. Tu,
\newblock {\em Differential Forms in Algebraic Topology},
\newblock Springer Verlag, New York, gtm 82 edition, 1982.

\bibitem{kirillov:87}
A.~N. {Kirillov} and N.~Y. {Reshetikhin},
\newblock {\em {Exact solution of the integrable XXZ Heisenberg model with
  arbitrary spin. I. The ground state and the excitation spectrum}},
\newblock J. Phys. A : Math. Gen. {\textbf{20}} (1987)   1565--1585.

\bibitem{fridkin:00}
V.~{Fridkin}, Y.~{Stroganov}  and D.~{Zagier},
\newblock {\em {Finite Size XXZ Spin Chain with Anisotropy Parameter $\Delta =
  1/2$}},
\newblock arXiv:nlin/0010021 (2000).

\bibitem{fridkin:00_2}
V.~{Fridkin}, Y.~{Stroganov}  and D.~{Zagier},
\newblock {\em {Ground state of the quantum symmetric finite-size XXZ spin
  chain with anisotropy parameter $\Delta = {\frac12}$}},
\newblock J. Phys. A: Math. Gen. {\textbf{33}} (2000)   L121--L125,
\newblock arXiv:hep-th/9912252.

\bibitem{fabricius:05}
K.~{Fabricius} and B.~M. {McCoy},
\newblock {\em {New Developments in the Eight Vertex Model II. Chains of Odd
  Length}},
\newblock J. Stat. Phys. {\textbf{120}} (2005)   37--70,
\newblock arXiv:cond-mat/0410113.

\bibitem{schwimmer:87}
A.~Schwimmer and N.~Seiberg,
\newblock {\em Comments on the n = 2,3,4 superconformal algebras in two
  dimensions},
\newblock Physics Letters B {\textbf{184}} {\textbf{2--3}} (1987)   191 -- 196.

\bibitem{dixon:88}
L.~Dixon, P.~Ginsparg  and J.~Harvey,
\newblock {\em {$\hat c=1$ Superconformal Field Theory}},
\newblock Nuclear Physics B {\textbf{306}} {\textbf{3}} (1988)   470 -- 496.

\bibitem{huijse:11_2}
L.~{Huijse},
\newblock {\em {Detailed analysis of the continuum limit of a supersymmetric
  lattice model in 1D}},
\newblock J. Stat. Mech. (2011)   P04004,
\newblock arXiv:1102.1700.

\bibitem{fendley:92}
P.~Fendley and K.~Intriligator,
\newblock {\em Scattering and thermodynamics of fractionally-charged
  supersymmetric solitions},
\newblock Nucl. Phys. B {\textbf{372}} (1992)   533.

\bibitem{fendley:92_1}
P.~{Fendley} and K.~{Intriligator},
\newblock {\em {Scattering and thermodynamics in integrable N = 2 theories}},
\newblock Nucl. Phys. B {\textbf{380}} (1992)   265--290,
\newblock arXiv:hep-th/9202011.

\bibitem{bajnok:04}
Z.~{Bajnok}, C.~{Dunning}, L.~{Palla}, G.~{Tak{\'a}cs}  and F.~{W{\'a}gner},
\newblock {\em {SUSY sine-Gordon theory as a perturbed conformal field theory
  and finite size effects}},
\newblock Nucl. Phys. B {\textbf{679}} (2004)   521--544,
\newblock hep-th/0309120.

\bibitem{hegedus:07}
{\'A}.~{Heged{\H u}s}, F.~{Ravanini}  and J.~{Suzuki},
\newblock {\em {Exact finite size spectrum in super sine-Gordon model}},
\newblock Nucl. Phys. B {\textbf{763}} (2007)   330--353.

\bibitem{bernard:90_2}
D.~Bernard and A.~LeClair,
\newblock {\em {The fractional supersymmetric sine-Gordon models}},
\newblock Phys. Lett. B {\textbf{247}} (1990)   309--316.

\bibitem{lerche:89}
W.~Lerche, C.~Vafa  and N.~P.~Warner,
\newblock {\em {Chiral rings in $N=2$ superconformal field theories}},
\newblock Nucl. Phys. B {\textbf{324}} (1989)   427--474.

\end{thebibliography}
\end{document}